\newif\ifsingle
\singletrue 

\newif\ifFullVersion
\FullVersiontrue 

\ifsingle
\documentclass[12pt,draftclsnofoot, onecolumn]{IEEEtran}		
\else		
\documentclass[10pt,final, twocolumn]{IEEEtran}
\fi

\usepackage{times}
\usepackage{amsmath,dsfont}
\usepackage{amssymb,amsthm}
\usepackage{epsfig,verbatim}
\usepackage{subcaption}
\usepackage{setspace}
\usepackage{color}
\usepackage{cite}
\usepackage{epstopdf}
\usepackage{graphics}
\usepackage{accents}
\usepackage{acronym}
\usepackage{booktabs}
\usepackage{mathtools} 
\usepackage{enumitem}
\usepackage{multirow}
\usepackage{dblfloatfix}
\usepackage[bookmarks,colorlinks]{hyperref}
\usepackage{gensymb}

\usepackage[ruled,linesnumbered]{algorithm2e}  

\SetKwInput{KwData}{\textbf{Init}} 
\let\oldnl\nl
\newcommand{\nonl}{\renewcommand{\nl}{\let\nl\oldnl}}

\newcommand{\myVec}[1]{{\boldsymbol{#1}}}
\newcommand{\myMat}[1]{{\boldsymbol{#1}}}
\newcommand{\mySet}[1]{\mathcal{#1}}
	
\newcommand{\mys}{{\myVec{s}}}	

\newcommand{\mypoint}{\myVec{s}}
\newcommand{\mydiff}{\myVec{d}}

\newcommand{\mydiffSet}{\mySet{D}}

\newcommand{\Weights}{\myVec{\varphi}}
\newcommand{\myCCP}{\myMat{P}^{\rm cc}}
\newcommand{\myCCPN}{\tilde{\myMat{P}}^{\rm cc}}
\newcommand{\myPTT}{\myMat{P}^{\rm tr}}

\newcommand{\Mem}{K}			 			
\newcommand{\Blklen}{B}			 			
\newcommand{\Blkset}{\mySet{\Blklen}}

\ifsingle

\newcommand{\figSpace}{\vspace{-0.2cm}}

\else

\newcommand{\figSpace}{\vspace{-0.2cm}}
\fi 

\acrodef{adc}[ADC]{analog-to-digital convertor}
\acrodef{cs}[CS]{compressed sensing}
\acrodef{dtft}[DTFT]{discrete-time Fourier transform}
\acrodef{dnn}[DNN]{deep neural network} 
\acrodef{csi}[CSI]{channel state information}
\acrodef{bpsk}[BPSK]{binary phase shift keying}
\acrodef{qpsk}[QPSK]{quadrature phase shift keying}
\acrodef{map}[MAP]{maximum a-posteriori probability}
\acrodef{snr}[SNR]{signal-to-noise ratio}
\acrodef{bs}[BS]{base station} 
\acrodef{iot}[IOT]{Interent of Things}
\acrodef{mimo}[MIMO]{multiple-input multiple-output}
\acrodef{siso}[SISO]{single-input single-output}
\acrodef{mse}[MSE]{mean-squared error}
\acrodef{pdf}[PDF]{probability density function}
\acrodef{rv}[RV]{random variable}
\acrodef{ml}[ML]{machine learning}
\acrodef{fec}[FEC]{forward error correction}
\acrodef{rs}[RS]{Reed-Solomon}
\acrodef{lti}[LTI]{linear time-invariant}
\acrodef{wss}[WSS]{wide-sense stationary}
\acrodef{psd}[PSD]{power spectral density}
\acrodef{ser}[SER]{symbol error rate} 
\acrodef{ber}[BER]{bit error rate} 
\acrodef{gd}[GD]{gradient descent}
\acrodef{sgd}[SGD]{stochastic gradient descent} 
\acrodef{isi}[ISI]{intersymbol interference}  
\acrodef{awgn}[AWGN]{additive zero-mean white real Gaussian noise} 
\acrodef{ut}[UT]{user terminal} 
\acrodef{mmw}[mmWave]{millimeter wave}
\acrodef{noma}[NOMA]{non-orthogonal multiple access}
\acrodef{mac}[MAC]{mulitple access channel}
\acrodef{fl}[FL]{Federated learning}
\acrodef{lstm}[LSTM]{long short-term memory}
\acrodef{maml}[MAML]{model-agnostic meta-learning}
\acrodef{sic}[SIC]{soft interference cancellation}
\acrodef{pmf}[PMF]{probability mass function}

\IEEEoverridecommandlockouts 

\title{Data Augmentation for Deep Receivers}
\author{  
	\IEEEauthorblockN{Tomer Raviv and Nir Shlezinger
	} 
	\thanks{
		Parts of this work were 
		presented at the 2022 IEEE Workshop on Signal Processing Advances in Wireless Communications (SPAWC) as the paper \cite{raviv2022adaptive}.	
		This project has received funding from the Israeli 5G-WIN consertium.
    The authors are with the School of ECE, Ben-Gurion University of the Negev, Beer-Sheva, Israel (e-mail: tomerraviv95@gmail.com, nirshl@bgu.ac.il).	}

	\vspace{-1.0cm}
	
}
\vspace{-0.75cm}

\begin{document}
	
	\maketitle

	\pagestyle{plain}
	\thispagestyle{plain}
	\begin{abstract} 
Deep neural networks (DNNs) allow digital receivers to learn to operate in complex environments. To do so, DNNs should  preferably be trained using large labeled data sets with a similar statistical relationship as the one under which they are to infer. For DNN-aided receivers, obtaining labeled data  conventionally involves pilot signalling at the cost of reduced spectral efficiency, typically resulting in access to limited data sets.
In this paper, we study how one can enrich a small set of labeled pilots data into a larger data set for training deep receivers. Motivated by the widespread use of data augmentation techniques for enriching visual and text data, we propose dedicated augmentation schemes that exploits the characteristics of digital communication data.  We identify the key considerations in data augmentations for deep receivers as the need for domain orientation, class (constellation) diversity, and low complexity. Following these guidelines, we devise three complementing augmentations that exploit the geometric properties of digital constellations. Our combined augmentation approach builds on the merits of these different augmentations to synthesize reliable data from a momentary channel distribution, to be used for training deep receivers. Furthermore, we exploit previous channel realizations to increase the reliability of the augmented samples.
The superiority of our approach is numerically evaluated for training several deep receiver architectures in different channel conditions. We consider both linear and non-linear synthetic channels, as well as the COST 2100 channel generator, for both single-input single-output and multiple-input multiple-output scenarios. We show that our combined augmentations approach allows DNN-aided receivers to achieve gains of up to $1$ dB in \acl{ber} and of up to $\times 3$ in spectral efficiency, compared to regular non-augmented training. 
Moreover, we demonstrate that our augmentations benefit training even as the number of pilots increases, and perform an ablation study on the different augmentations, which shows that the combined approach surpasses each individual augmentation technique.
\end{abstract}
	\vspace{-0.4cm}
	\section{Introduction}
\vspace{-0.1cm} 
	
The emergence of deep learning has notably impacted numerous applications in various disciplines. 
The ability of \acp{dnn} to learn to efficiently carry out complex tasks from data has spurred a growing interest in their usage for receiver design in digital communications \cite{gunduz2019machine,simeone2018very,balatsoukas2019deep, farsad2020data}. While traditional receiver algorithms are channel-model-based, relying on mathematical modeling of the signal transmission, propagation, and reception, \ac{dnn}-based receivers have the potential to operate efficiently in model-deficient scenarios where the  channel model is unknown, highly complex, or difficult to optimize for~\cite{farsad2020data}, and can thus greatly contribute to meeting the constantly growing requirements of wireless systems in terms of throughput, coverage, and robustness \cite{saad2019vision}. Generally, deep learning can be integrated with receiver design either by using conventional black-box DNN architectures trained end-to-end; or by leveraging model-based solutions \cite{shlezinger2022model,shlezinger2020model,eldar2022machine,cammerer2020trainable}, whereby specific blocks of a receiver's architecture are replaced  by neural networks, e.g., via deep unfolding \cite{balatsoukas2019deep}. 

While deep learning thrives in domains such as computer vision and natural language processing, many challenges still limit its widespread applicability in digital communications. One of these challenges arises from the fact that \acp{dnn} rely on  (typically labeled) data sets to learn their mappings, and are thus usually applied when the  function being learned is stationary for a long period of time, such that one can collect sufficient data for learning. For example, data sets for machine translation tasks, which are nowadays dominated by deep learning systems, are usually composed of a sentence in an origin language, and its respective translation in a destination language, which do not change much with time. One can thus collect a large volume of such pairs for training, obtaining a huge data set for the task at hand \cite{roh2019survey}. 

In contrast, the operation of digital receivers is rarely stationary, as communication channels (and thus the data distribution and in turn the system task) change considerably over time. Moreover, \acp{dnn} are  highly-parameterized models, and their training requires massive data sets. In  digital receivers, labeled data corresponding to the current channel is often obtained from pilots, whose transmission degrades the spectral efficiency. Thus, deployed digital receivers are likely to have access to small labeled data sets corresponding to the current task, hardly in the scale needed to train \acp{dnn}. 
Consequently, a core challenge associated with the usage of \ac{dnn}-aided receivers stems from the need to aggregate sufficient task-specific data, i.e., data corresponding to the instantaneous setting where the receiver is required to operate.


Various strategies have been proposed  to facilitate the application of \acp{dnn}-aided receivers with limited data sets. 
The first approach trains with data obtained from simulations and past measurements, while using the limited data corresponding to the instantaneous channel to estimate some missing parameters, e.g., a channel matrix. These parameters are then used by the network, either as an input \cite{honkala2021deeprx,zhao2021deep} or as part of some internal processing  \cite{he2020model,khani2020adaptive, samuel2019learning, goutay2021machine,pratik2020re,schmid2022neural,baumgartner2022soft}. 
This  involves imposing a relatively simple model on the channel, typically a linear Gaussian model, which limits their suitability in the presence of complex channel models.  

An alternative approach designs receivers with compact task-specific \acp{dnn} such that they can be trained with relatively small data sets, by integrating them into classic receiver processing via model-based deep learning techniques \cite{shlezinger2020model, shlezinger2022model}, see, e.g., \cite{shlezinger2020data,shlezinger2019deepSIC, shlezinger2019viterbinet,raviv2020data,van2022deep}.
These hybrid model-based/data-driven receivers build on existing communication models and incorporate deep learning into their design. The resulting architectures are agnostic to the  channel model and its parameters, imitating the operation of a detection algorithm with no channel knowledge. Although these \ac{dnn}-aided receivers are compact in terms of trainable parameters, the amount of data required in training is still likely to be larger than the amount corresponding to the instantaneous channel one can typically obtain in real-time.

A third approach aims to optimize the training algorithm itself, rather than the architecture, facilitating rapid convergence of the training procedure with limited data.  Optimizing the hyperparameters that govern the optimization process, e.g., the initialization of the trainable weights, can intuitively be seen as a method to share data from different time steps, increasing the effective labeled data sets.
This can be achieved by leveraging  data corresponding to past channel realizations in order to tune the optimization hyperparameters via Bayesian optimization techniques~\cite{wang2021jointly}; model-agnostic meta-learning  \cite{park2020learning,park2020meta,simeone2020learning}; and predictive meta-learning for learning transition patterns in time-varying channels \cite{raviv2021meta,raviv2022online}. 

The aforementioned strategies focus on either the architecture or the training algorithm. A complementary approach to cope with limited data sets, which can be combined with any of the above techniques, is to generate more data at the receiver. Current techniques for obtaining additional labeled data in wireless communications aim at assigning labels to channel outputs not associated with pilots by exploiting the presence of channel coding for self-supervision \cite{shlezinger2019viterbinet, shlezinger2020inference, shlezinger2019deepSIC,teng2020syndrome,schibisch2018online} or by using active learning techniques \cite{be2019active, finish2022symbol}. However, these approaches cannot synthesize more data than the block length (typically generating much less \cite{finish2022symbol}), and the resulting data set may be too small for training.

A common practice in the deep learning literature  is to synthesize data samples via {\em data augmentation}. Data augmentation is a set of techniques that enrich a trainable model with new unseen synthetic data, where each technique builds on partial knowledge of the data  characteristics~\cite{shorten2019survey,nalepa2019data,feng2021survey,park2019specaugment}, typically exploiting some underlying translation invariance.  For instance, in image classification, one can use a labeled image to generate multiple different images with the same label by, e.g., rotating or clipping it~\cite{shorten2019survey,nalepa2019data}. While data augmentation is a widely used deep learning technique, it is highly geared towards image and language data. This motivates the study of augmentation techniques for enriching data in digital communications, as means to facilitate the training of \ac{dnn}-aided receivers with limited labeled data sets.

\subsection*{Main Contributions}

In this paper we propose a data augmentation framework for \ac{dnn}-aided receivers. Our proposed method introduces a set of data synthesis techniques which exploit expected structures of digital communication symbols and their corresponding channel outputs to enrich a small labeled data set with many reliable samples. The resulting enriched data can then be used for training, so as to reduce the error rate obtained by the receiver, and specifically the epistemic uncertainty that exists due to limitations of available training data \cite{jose2022information}. 
Our main contributions are summarized as follows: 
\begin{itemize}
    \item \textbf{Proposed framework:} We formulate a communication-directed data augmentations framework, which embodies the specific characteristics of the digital communication data used for training \ac{dnn}-aided receivers.
    \item \textbf{Communications-tailored augmentations:} We propose several augmentations that exploit innate traits of digital constellations. We exploit both expected symmetries in channel outputs as well as identified communication-specific translation invariance properties to generate new synthetic samples for training. Our proposed augmentation techniques are complementary of each other, and are combined into a unified method.
    \item \textbf{Adaptive augmentations for time-varying channels:} We extend our augmentation techniques to temporally adapt for deep receivers operating in block-fading channels. This allows leveraging past channels to generate additional synthetic data to be used for online training in time-varying conditions. 
    \item \textbf{Extensive experimentation:} We extensively evaluate the proposed training scheme for training various \ac{dnn}-aided receiver architectures. We consider different channel profiles for both multipath \ac{siso} channels as well as memoryless \ac{mimo} systems. We demonstrate consistent benefits by using our approach, which amount to gains of up to $1$ dB in \acl{ber}, and of up to $\times 3$ in spectral efficiency, compared with non-augmented training. Moreover, we show that our augmentations benefit training even as the number of pilots increases, and perform an ablation study on the different augmentations, which shows that combining the augmentation techniques benefits upon using each of the individual methods.
\end{itemize}

 The rest of this paper is organized as follows:  Section~\ref{sec:system_model} details the system model and the operation of \ac{dnn}-aided receivers. Section~\ref{sec:data_augmentations_static} formulates the framework and presents the augmentation techniques for static channels, where a single small data set corresponding to a specific channel realization is to be synthetically enriched for training. Section~\ref{sec:data_augmentations_dynamic} extends the augmentation techniques to dynamic communication setups, accounting for continuous variations in block-fading channels. Experimental results and concluding remarks are detailed in Section~\ref{sec:numerical_evaluations} and Section~\ref{sec:conclusion}, respectively.

Throughout the paper, we use boldface letters for vectors, e.g., ${\myVec{x}}$; $({\myVec{x}})_i$ denotes
the $i$th element of ${\myVec{x}}$. Upper-cased boldface letters denote matrices, e.g., $\myMat{X}$, with $\myMat{I}_n$ being the $n\times n$ identity matrix. 
Calligraphic letters, such as $\mySet{X}$, are used for sets, with $|\mySet{X}|$ being the cardinality of $\mySet{X}$. We denote by  $\mySet{R}$ and $\mySet{C}$ the sets of real and complex numbers, respectively, while $\mySet{N}(\cdot,\cdot) $ is the Gaussian distribution. The operation $(\cdot)^H$ stands for the conjugate transpose operation.

	\vspace{-0.2cm}
	\section{System Model}
\label{sec:system_model}
\vspace{-0.1cm}

In the section, present the system model. To that aim, we first describe the communication model in Subsection~\ref{subsec:communication_model}. Then, we discuss the receiver processing and the considered problem formulation of data augmentation for \ac{dnn}-aided receivers in Subsections~\ref{subsec:Receiver} and \ref{subsec:Problem}, respectively.

\vspace{-0.2cm}
\subsection{Communication Model}
\label{subsec:communication_model}
\vspace{-0.1cm} 
We consider a digital communication system in discrete-time. 
Let $\myVec{s}_i\in\mySet{S}$ be a symbol transmitted from constellation $\mySet{S}$ (with its size denoted $|\mySet{S}|$) at the $i$th time instance within a block of $\Blklen ^{\rm tran}$ symbols, i.e., $i \in \{1,2,\ldots, \Blklen ^{\rm tran}\} = \Blkset$. The transmitted symbols block $\mys ^{\rm tran}:= \{\myVec{s}_i\}_{i\in \Blkset}$   is divided into $\Blklen^{\rm pilot}$ pilots that are known to the receiver and appear first,  denoted $\myVec{s}^{\rm pilot}$, and  $\Blklen^{\rm info} = \Blklen^{\rm tran}-\Blklen^{\rm pilot}$ information symbols, denoted $\myVec{s}^{\rm info}$, that contain the digital message conveyed to the receiver. The channel output at time instance $i$ is denoted by $\myVec{y}_i$, which takes values in the set $\mySet{Y}$, and the received block is $\myVec{y}^{\rm rec}:= \{\myVec{y}_i\}_{i\in \Blkset}$. Similarly as the channel input, the channel output can be separated by the receiver into its pilot and information parts denoted $\myVec{y}^{\rm pilot}$ and $\myVec{y}^{\rm info}$, respectively. We assume that the channel, i.e., the mapping from $\myVec{s}$ into $\myVec{y}$, is constant within the block of $\Blklen^{\rm tran}$ channel uses, which thus corresponds to the coherence duration of the channel.

{\bf Channel Models:} 
While the above channel model is generic, we focus mostly on the common setting of communication over casual finite-memory channels. 
Such models accommodate a broad range of communication scenarios, including the following settings which are considered in our experimental study:
\begin{enumerate}
    \item \emph{Finite-Memory \ac{siso} Channels}: In multipath \ac{siso} channels, the channel output $\myVec{y}_{i}$ (which is scalar and thus written as $y_i$) is given by a stochastic function of the last $\Mem > 0$ transmitted symbols, which can  be stacked into the $\Mem\times 1$ vector $\myVec{s}_i$, where $\Mem$ is the memory length. Letting $M$ be the alphabet size, the number of different constellation combinations that $\myVec{s}_i$ can represent is  $|\mySet{S}|=M^\Mem$. Independence between different blocks is obtained by setting a guard interval of at least $\Mem-1$ time instances prior to the beginning of the block.
    \item \emph{Flat \ac{mimo} Channels}: Another channel of interest is the memoryless uplink \ac{mimo} channel, where $K$ single-antenna transmitters (users) communicate with a receiver equipped with $N$ antennas. Again letting $M$ be the alphabet size, the channel input is the $K\times 1$ vector $\myVec{s}_i$, which can take  $|\mySet{S}|=M^K$ different values, while the channel output is an $N\times 1$ vector. 
\end{enumerate}
Based on the above examples, we henceforth consider the symbol space $\mySet{S}$ to be comprised of $K \times 1$, i.e., $\mySet{S}\subset \mySet{C}^K$, and the channel outputs $\myVec{y}_i$ to be $N\times 1$ vectors, for some fixed $N,K>0$. 

\vspace{-0.2cm}
\subsection{Receiver Operation}
\label{subsec:Receiver}
\vspace{-0.1cm} 
The receiver processing is aided by a  \ac{dnn} which relies on labeled data to learn its mapping. In particular, the receiver recovers the information symbols from $\myVec{y}^{\rm info}$ using a \ac{dnn}-aided detection mapping, which yields the estimate $\hat{\myVec{s}}(\myVec{y}^{\rm info}; \Weights)$, where  $\Weights$  are the  parameters of the \ac{dnn}.
The parameters $\Weights$ are obtained by training, using the available labeled data, denoted by $\mySet{Q}^*$, which corresponds to the underlying channel and is comprised of pairs of channel outputs $\myVec{y}_i$ and their corresponding symbol vectors $\myVec{s}_i$.
The training module follows a (stochastic) gradient-based optimization with empirical loss measure $\mySet{L}_{\mySet{Q}^*}$, e.g., $I_{\rm sgd}$ iterations of the form
 \begin{align}
\label{eq:local_update} 
\Weights^{(t+1)} = \Weights^{(t)} - \eta \hat{\nabla}_{\Weights}\mySet{L}_{\mySet{Q}^*}(\Weights^{(t)}),
 \end{align}
 where $\eta > 0$ is the learning rate and $\hat{\nabla}_{\Weights}\mySet{L}$ is a stochastic approximation of the gradient of the empirical loss with respect to the trainable parameters. Since symbol detection is a classification task, we focus on $\mySet{L}_{\mySet{Q}^*}$ being the empirical cross-entropy loss, such that \eqref{eq:local_update} approximates the solution to
 \begin{align}
\label{eq:meta_1}
  \mathop{\arg \min}_{\Weights} \Big\{ \mySet{L}_{\mySet{Q}^*}(\Weights) = -\sum_{(\myVec{y}_i, \myVec{s}_i) \in\mySet{Q}^* }\log \hat{P}_{\Weights}\left(\myVec{s}_{i}|\myVec{y}_{i} \right)\Big\},
\end{align}
with $\hat{P}_{\Weights}(\cdot|\cdot)$ representing the soft (probabilistic) estimates produced by the receiver with \ac{dnn} parameters $\Weights$.

\begin{figure}
    \centering
    \includegraphics[width=\columnwidth]{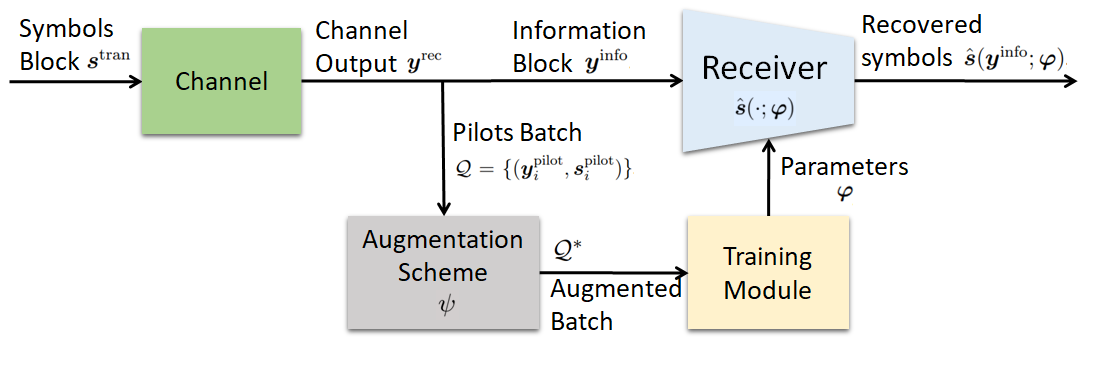}
    \caption{\ac{dnn}-aided receiver operation illustration, with the block marked with $\psi$ representing the augmentation scheme.}
    \label{fig:transmission}
\end{figure}

While the \ac{dnn}-aided receiver trains with a labeled data set for which the transmitted symbols are known, its performance is measured in its ability to recover the information data from the known constellation $\mySet{S}$. Consequently, the considered performance measure for evaluating the parameters $\Weights$ trained via \eqref{eq:local_update} is the \ac{ber} over the information block, given by 
\begin{equation}
\label{eqn:ErrorRate}
	e(\Weights) = \frac{1}{\Blklen^{\rm info}}\sum_{i= \Blklen^{\rm pilot} + 1}^{\Blklen^{\rm tran}} \Pr\left( \hat{\myVec{s}}_i(\myVec{y}^{\rm info};\Weights) \neq  \myVec{s}_i^{\rm info} \right). 
\end{equation} 
An illustration of the system is depicted in Fig.~\ref{fig:transmission}. 

\vspace{-0.2cm}
\subsection{Problem Formulation}
\label{subsec:Problem}
\vspace{-0.1cm}  
The labeled data set corresponding to the current channel to which the receiver has access is given by  $\mySet{Q}= \{(\myVec{y}_i^{\rm pilot},\myVec{s}_i^{\rm pilot})|i \in \{1,2,\ldots, \Blklen ^{\rm pilot}\} \}$, i.e., the known pilots and their respective received values.  When the number of samples $B^{\rm pilot}$ is relatively small, training via \eqref{eq:local_update} with the data set $\mySet{Q}^* = \mySet{Q}$ is likely to result in poor information error rates. Consequently, we opt to augment new data for training. 

Our goal is to derive an augmentation scheme $\psi:\mySet{Q} \mapsto \mySet{Q}^*$ that enriches the available data $\mySet{Q}$ into the training set $\mySet{Q}^*$. In particular, after training with $\mySet{Q}^*$, one should result in parameters $\Weights$ that improve upon training with $\mySet{Q}$ in the sense of the \ac{ber} objective   \eqref{eqn:ErrorRate}. The integration of data augmentation into the overall receiver operation is illustrated in the bottom part of Fig.~\ref{fig:transmission}.

 
	\vspace{-0.2cm}
	\section{Data Augmentations for Static Channels}
	\label{sec:data_augmentations_static}
	\vspace{-0.1cm}
Here, we present the proposed data augmentation techniques, considering a static setting, in which one has access to a small data set corresponding to a single channel realization where it should also infer.
In Subsection~\ref{subsec:main_pillars} we introduce the rationale for using augmentations, and discuss the main pillars for data augmentation techniques in light of the operation of \ac{dnn}-aided communication receivers. We follow up with the communication-oriented augmentations in  Subsection~\ref{subsec:augmentations_static}, and provide a discussion in  Subsection~\ref{subsec:discussion}.  

\vspace{-0.4cm}
\subsection{Rationale}
\label{subsec:main_pillars}

Data augmentation is a widely-spread practice in machine learning. 
For example, in speech processing, data augmentation includes several domain-tailored transforms such as time warping and frequency masking \cite{park2019specaugment}, which use observed sequences to generate new sequences that the system should learn to cope with. In general, the purpose of using augmentations is to enrich a data set by exploiting domain knowledge regarding the nature of the data, and by doing so, encourage the trainable model to learn some known property of the data. The enriched set is then used for training the model with samples that may be observed during inference, but are not included in the original training set.

In digital communications, one is often faced with scarce data scenarios corresponding to a given task, and thus augmentations can aid in obtaining new data for training \ac{dnn}-aided receivers. However, existing data augmentation techniques are highly geared towards structures and invariances exhibited by data in conventional machine learning domains such as visual, language, and speech data, and thus cannot be directly implemented in digital communications. 
Therefore, 
in the following sections, we develop augmentation techniques for data used by digital receivers.  Our approach accounts for the following data augmentation considerations:
\begin{enumerate}[label={\em C\arabic*}]
    \item \textbf{Domain-oriented} - \label{itm:domain} Data augmentations are inherently domain-oriented \cite{nooraiepour2021hybrid}. In our case, we exploit the inherent structure of digital constellations, e.g., that some specific rotations applied to constellation symbols yield other valid symbols, to formulate constellation-preserving transformations. 
    \item \textbf{Diversity} - \label{itm:diversity}  The number of classes grows quickly with the system parameters such as the constellation size and the number of users. As such, the number of samples per class should be kept above some minimal threshold to avoid class imbalances~ \cite{yang2022image,be2019active,finish2022symbol}. Our devised augmentations should suffice this threshold. 
    \item \textbf{Complexity} - \label{itm:complexity} As our proposed data augmentation is to be implemented by a digital receiver in real-time, we must keep the sample-complexity low and the computation lightweight. This limits the usage of data augmentation approaches based on additional generative \acp{dnn}, commonly considered in conventional deep learning domains\footnote{For example, in \cite{belousov2021mobilestylegan}, the authors note that \textit{"The high computational complexity (of generative models) makes it difficult to deploy state-of-the-art generative models to edge devices."} } \cite{almahairi2018augmented,laskin2020reinforcement}, and motivates us to focus on simpler augmentation methods based on geometric considerations and class translations.  
\end{enumerate}

\begin{figure*}  
\begin{subfigure}{0.45\textwidth}
\includegraphics[width=\linewidth]{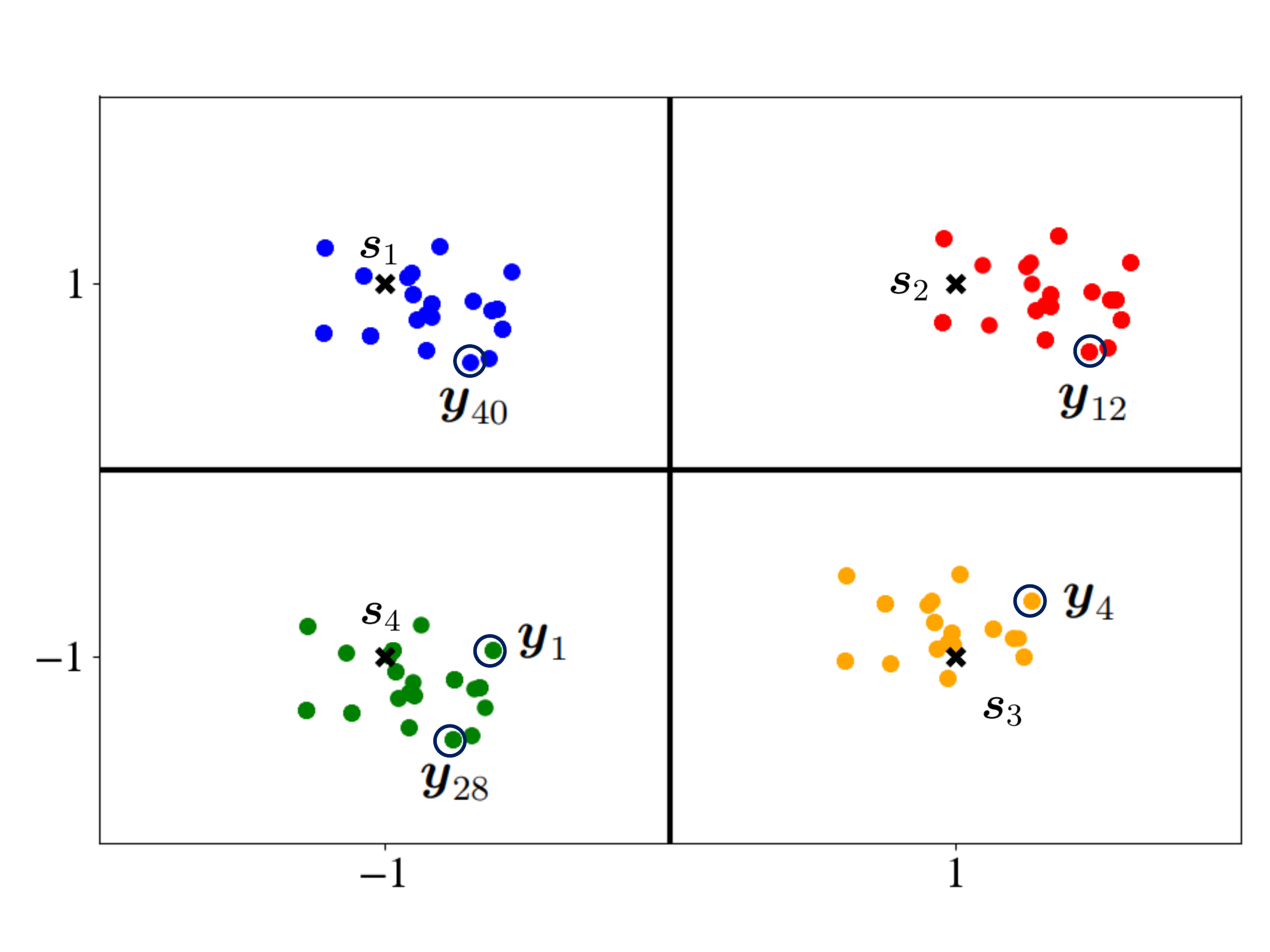}
\caption{Constellation points and received pilots}
\label{fig:constellation}
\end{subfigure}
\hfill 
\begin{subfigure}{0.45\textwidth}
\includegraphics[width=\linewidth]{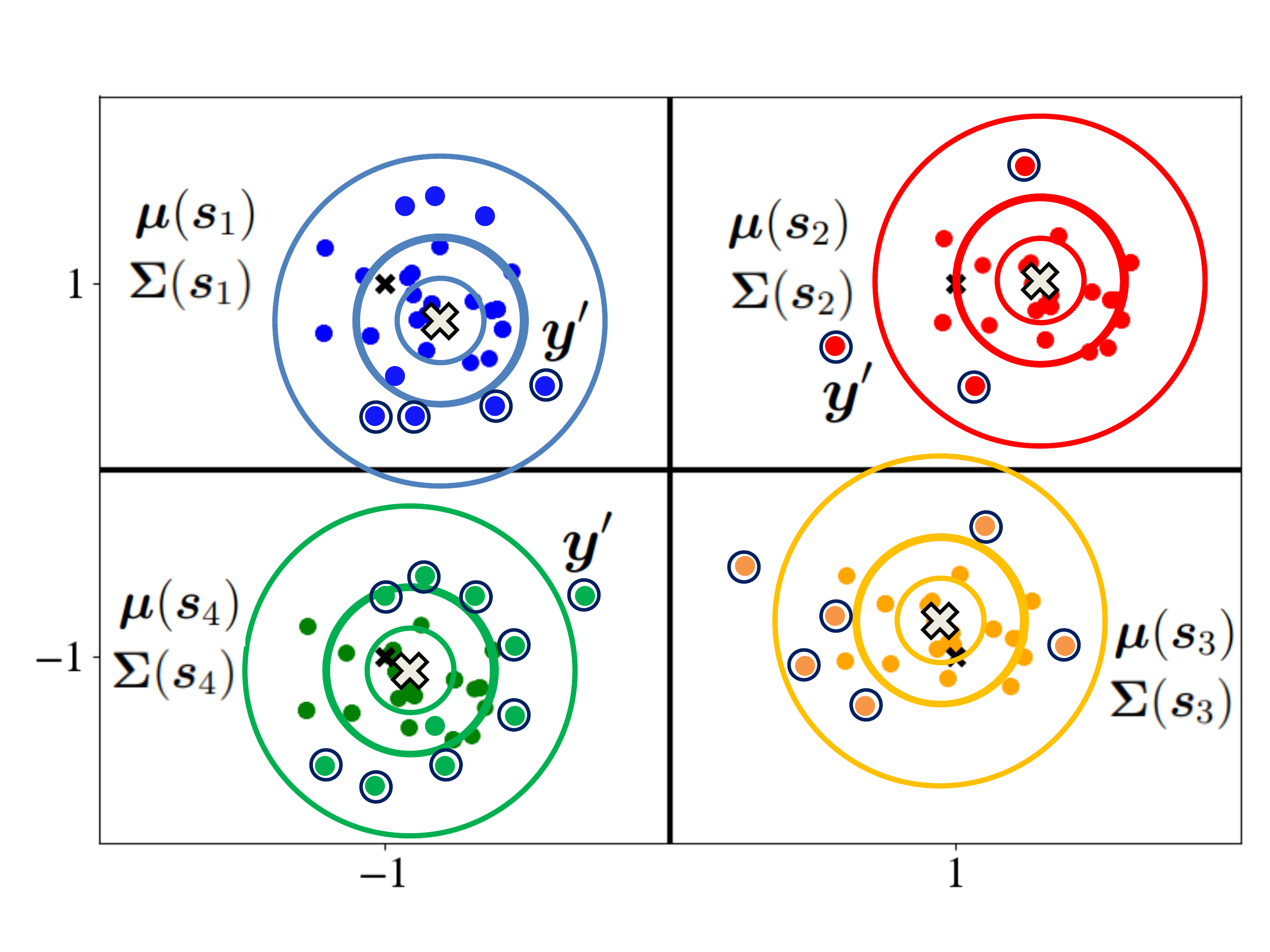}
\caption{Geometric augmentation}
\label{fig:geomtric}
\end{subfigure}

\bigskip  

\begin{subfigure}{0.45\textwidth}
\includegraphics[width=\linewidth]{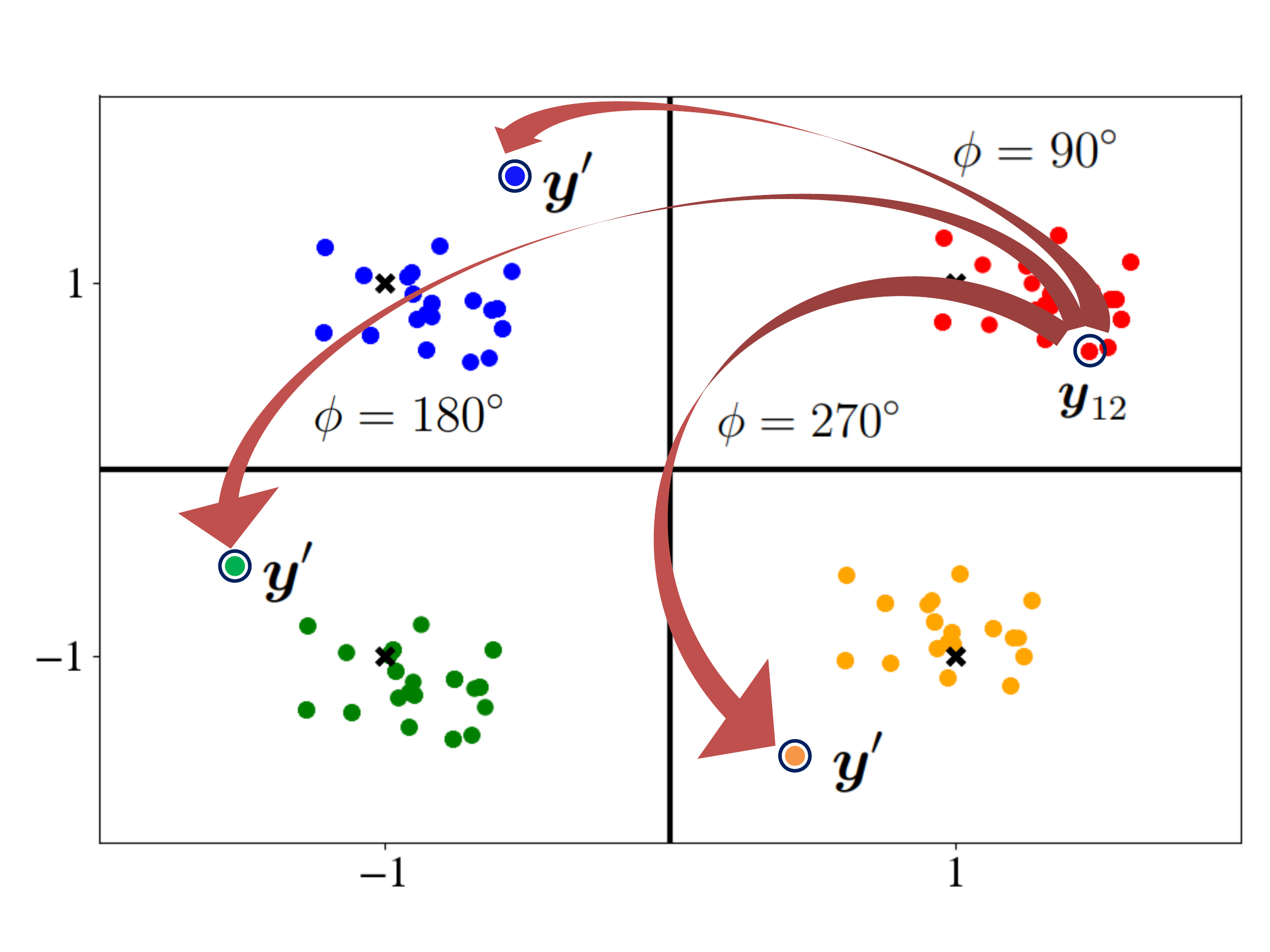}
\caption{Constellation-conserving rotation}
\label{fig:rotation}
\end{subfigure}
\hfill 
\begin{subfigure}{0.45\textwidth}
\includegraphics[width=\linewidth]{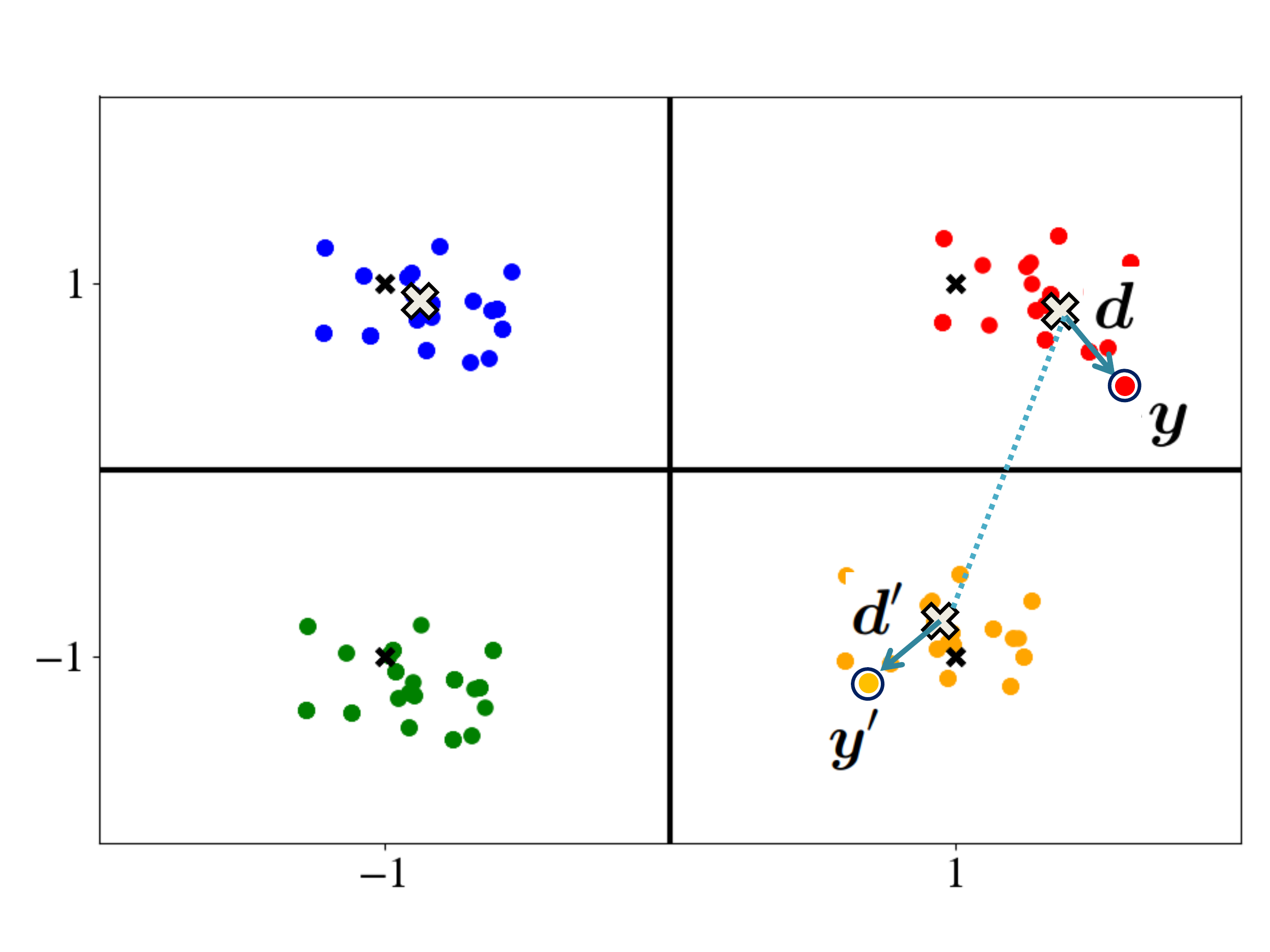}
\caption{Translation augmentation}
\label{fig:translation}
\end{subfigure}

\caption{Augmentations exemplified on QPSK constellation} 
\end{figure*}

 
\vspace{-0.4cm}
\subsection{Augmentations}
\label{subsec:augmentations_static}

Designing data augmentations requires one to identify label-invariant transformations that account for some knowledge about the statistical model of data. For \ac{dnn}-aided digital receivers, this statistical model is the underlying input-output relationship induced by the communication channel. In the following, we detail three different and complementary augmentations: $(1)$ geometric augmentation; $(2)$ constellation-preserving projection; and $(3)$ translation augmentation. Each of these methods  originates from some expected behaviour of wireless communications channels.

\subsubsection{Geometric Augmentation} 
Our first augmentation technique draws inspiration from the common model of wireless channels as being decomposable into a channel response and an additive noise term. In such cases, one can write
\begin{equation}
    \label{eqn:addNoiseChannel}
    \myVec{y}=g(\myVec{s})+\myVec{n},
\end{equation}
where $g$ is the channel input-to-output transformation, and $\myVec{n}$ is an additive distortion term which is independent of $\myVec{s}$, e.g., thermal noise and interference.  This representation indicates that channel outputs corresponding to the same input tend to form clusters, as illustrated in Fig.~\ref{fig:constellation}, where the intra-cluster distribution is invariant of the center, since the noise is independent of the  signal. 
This geometrical interpretation of the by-class clusters can be employed to generate additional class-specific data.

In particular, we propose to use the existing labeled data $\mySet{Q}$ to identify clusters in the observations space from which the received symbols are likely to originate. Our method is based on clustering the channel outputs in $\mySet{Q}$ and setting $\myVec{\mu}$  and $\myMat{\Sigma}$ to be the cluster-wise centers and covariances, respectively.  
%
Using the identified clusters, we suggest to synthesize data corresponding to a constellation point $\mypoint$ from a conditional Gaussian distribution $\mathcal{N}(\myVec{\mu}(\mypoint),\,\myMat{\Sigma}(\mypoint))$. 

Therefore, in order to enrich a labeled data set $\mySet{Q}$ into $\mySet{Q}^*$, we use $\mySet{Q}$ to determine $\myVec{\mu}$  and $\myMat{\Sigma}$ for each $\mypoint\in\mySet{S}$. To that aim, we define the index set corresponding to $\mypoint$ as 
\begin{equation}
    \label{eqn:IdxSet}
    \mySet{I}_{\mySet{Q}}(\mypoint) := \{i| \mypoint_i \in \mySet{Q}; \mypoint_i =  \mypoint \}.
\end{equation}
We can now  estimate the moments of the clusters as
 \begin{equation} \label{eq:mean_calculation}
    \myVec{\mu}(\mypoint) =\frac{1}{|\mySet{I}_{\mySet{Q}}(\mypoint)|} \sum_{\myVec{y}_i \in \mySet{Q}: i \in  \mySet{I}_{\mySet{Q}}(\mypoint)} \myVec{y}_{i},
\end{equation}
and similarly 
 \begin{equation} \label{eq:variance_calculation}
    \!\!\myMat{\Sigma}(\mypoint) \!=\!\frac{1}{|\mySet{I}_{\mySet{Q}}(\mypoint)|} \sum_{\myVec{y}_i \in \mySet{Q}: i \in  \mySet{I}_{\mySet{Q}}(\mypoint)} \!\!\!\!(\myVec{y}_{i}\! - \!\myVec{\mu}(\mypoint))(\myVec{y}_{i} \!-\! \myVec{\mu}(\mypoint))^H.
\end{equation}
To avoid empty sets $\mySet{I}(\mypoint)$, the pilot sequences should preferably be designed such that each class has at least a single representative. The augmentation is depicted in Fig.~\ref{fig:geomtric}.
Once the clusters are estimated, they are used to draw synthetic samples $\myVec{y}^{\prime}$ for some corresponding label $\mypoint^{\prime}$. We generate these synthetic samples from the conditional distribution $\mathcal{N}(\myVec{\mu}(\mypoint),\,\myMat{\Sigma}(\mypoint))$. 

\subsubsection{Constellation-Conserving Projections} 
Many digital constellations, including phase shift keying and quadrature amplitude modulation, are symmetric by design. This means that there exists a discrete set of linear projections $\{{\myCCP}\}$ which are \textit{constellation-conserving}. A matrix $\myMat{P}^{\rm cc}$, with size of {$K \times K$}, represents a constellation-conserving projection if for every $\myVec{s}\in\mySet{S}$, it is satisfied that $\myCCP\myVec{s}\in\mySet{S}$ as well. This property may be exploited to generate additional training data when the underlying channel meets the following two requirements:
\begin{itemize}
    \item There exists an $N\times N$ matrix $\myCCPN$ such that the  channel mapping $g(\cdot)$ in \eqref{eqn:addNoiseChannel} (approximately) satisfies $ \myCCPN g(\myVec{s}) \approx  g(\myCCP \myVec{s})$.
    \item The distribution of the noise $\myVec{n}$ in \eqref{eqn:addNoiseChannel} is invariant to the transformation, i.e., that $\myVec{n}^{\rm cc} = \myCCPN\myVec{n}$ obeys (approximately) the same distribution as  $\myVec{n}$.
    %
\end{itemize}
When the above hold, one can use a labeled pair $(\myVec{y}, \myVec{s})$ to generate additional data $(\myVec{y}^\prime, \myVec{s}^\prime)$ by rotating the channel output. This follows since there exists $\myVec{s}^\prime \in \mySet{S}$ such that
\begin{align}
\myVec{y}^\prime &=\myCCPN\myVec{y} = \myCCPN g(\myVec{s}) + \myCCPN\myVec{n} \notag \\
&\approx  g(\myCCP\myVec{s}) + \myVec{n}^{\rm cc} = g(\myVec{s}^\prime) + \myVec{n}^{\rm cc}. \label{eqn:constpres1} 
\end{align}
The joint distribution of  $\myVec{s}^\prime = \myCCP \myVec{s}$ and $\myVec{y}^\prime$ in \eqref{eqn:constpres1} is (approximately) identical to that of $(\myVec{y}, \myVec{s})$.

 
To design constellation-preserving augmentations, we note that communication channel mappings are often commutative to scalar multiplication, i.e., $g(a \myVec{s}) = a g(\myVec{s})$ for scalar $a$. This holds for instance, when the channel is linear or piecewise linear. Consequently, we suggest to utilize discrete rotations transformations, for which $\myCCP = e^{j \phi} \myMat{I}_K$ and $\myCCPN = e^{j \phi} \myMat{I}_N$ where $\phi\in[0,2\pi)$. For example, under a  \ac{qpsk} constellation ($M=4$), one can set $\phi = \frac{\pi m}{2}$ for any $m \in \{0,1,2,3\}$ such that $\myCCP \myVec{s}$ is a valid constellation point. Furthermore, the distribution of noise signals is often well-approximated as being rotation-invariant, as is the case for, e.g., circularly-symmetric Gaussian $\myVec{n}$. Consequently, applying these permutations is likely to be constellation-preserving and satisfy the above requirements. 

The resulting proposed augmentation operates by randomly choosing $\myCCP$ out of the set of constellation preserving rotations $\{e^{j \phi} \myMat{I}\}$, and applying it to a labeled sample in $\mySet{Q}$: 
\begin{equation} \label{eq:constellation_conserving_rotation}
     \myVec{y}^{\prime} = \myCCPN \myVec{y},\quad 
     \mypoint^{\prime} = \myCCP \mypoint.
\end{equation}
The augmented pair $(\myVec{y}^{\prime},\mypoint^{\prime})$ is added to the buffer $\mySet{Q}^*$. An illustration of the generation of such a synthetic example is depicted in Fig.~\ref{fig:rotation}.


\subsubsection{Translation Augmentation} 
We further exploit the idea of projections to share and exploit data from different clusters, by \textit{translation} of samples across the clusters. Here, we again build upon the additive distortion model in \eqref{eqn:addNoiseChannel}, but instead of generating new samples around a given cluster in a random fashion, we synthesize a new {\em realization} of the distortion, which is used together with the cluster center to create a new observation.

To formulate this idea, let $\myVec{\mu}(\mypoint)$ be the cluster center estimated in \eqref{eq:mean_calculation},  and $\mydiff$ be the empirical difference vector estimated as 
\begin{equation} \label{eq:difference_calculation}
    \mydiff =\myVec{y} - \myVec{\mu}(\mypoint).
\end{equation}
We aim to translate a known realization $\mydiff$ across different clusters via a translation operation defined via the $N\times N$ translation matrix $\myPTT$, such that $\mydiff^{\prime} = \myPTT \mydiff$.
Specifically, to generate a labeled augmented pair $(\myVec{y}^{\prime},\mypoint^{\prime})$ from a given $(\myVec{y}, \mypoint)$, we first compute $\mydiff$ via \eqref{eq:difference_calculation}. Then, we select $\myVec{s}^\prime \neq \myVec{s}$, and generate a synthetic channel output via:
\begin{align}
\myVec{y}^\prime &= \myVec{\mu}(\mypoint^{\prime}) + \mydiff^{\prime} = \myVec{\mu}(\mypoint^{\prime}) + \myPTT\mydiff \notag \\ 
&= \myPTT\myVec{y} + \myVec{\mu}(\mypoint^{\prime}) - \myPTT\myVec{\mu}(\mypoint). \label{eqn:translation1} 
\end{align}
Now, by introducing the notation $\Delta = \myVec{\mu}(\mypoint^{\prime}) - \myPTT\myVec{\mu}(\mypoint)$, we obtain the synthetic channel output $\myVec{y}^\prime$ as
\begin{equation}
\myVec{y}^\prime = \myPTT\myVec{y} + \Delta \label{eqn:translation2} 
\end{equation}
and the augmented pair $(\myVec{y}^{\prime},\mypoint^{\prime})$ is added to the buffer $\mySet{Q}^*$. 

The selection of the translation matrix $\myPTT$ captures the differences between the distortion realizations added for different symbols. Since the magnitude of the distortion is often invariant to the transmitted symbols, we opt unitary translation matrices. In particular, for the sake of simplicity, we employ discrete rotations transformations here as in the constellation-conserving projections case, for which $\myPTT = e^{j \phi} \myMat{I}_N$ where $\phi\in[0,2\pi)$. For example, under \ac{qpsk} constellation ($M=4$), one can set $\phi = \frac{\pi m}{2}$ for any $m \in \{0,1,2,3\}$.

As opposed to the geometric augmentation approach described above, translation augmentation does not impose a statistical model on the distortion for generating new samples. Instead, it re-uses observed realizations of the distortion term of a sample in some class for augmenting samples in other classes. 

\subsubsection{Combined Scheme} 
Each of the aforementioned methods allows to generate additional new data from an available labeled set comprised of channel outputs and their corresponding transmitted symbols. As these methods are complementary of each other, we combine them into an overall data augmentation method  by applying them subsequently.  

First, we initialize the buffer containing the data to be used for training using the available limited set, i.e., by setting $\mySet{Q}^*=\mySet{Q}$. To realize the diversity consideration \ref{itm:diversity}, we augment each sample into $\kappa$ new samples using any combination of the above techniques. All synthetic samples are added to the augmented training batch $\mySet{Q}^*$. When all three augmentations are used, this results in a set of $|\mySet{Q}^*| = (3\kappa+1) \cdot |\mySet{Q}|$ labeled samples to be used for training $\Weights$, as illustrated in Fig.~\ref{fig:transmission}. Note the each augmentation may be applied on its own, or in consecutive pairs. However, we provide empirical demonstration in Subsection~\ref{subsec:ablation_study} that augmented pairs which are the output of two different augmentations are often approximately independent of each other, causing the gain from each augmentation to sum up. The resulting combined augmentation policy $\psi_{\rm static}:\mySet{Q} \mapsto \mySet{Q}^*$ for static channels, combining all of the aforementioned techniques sequentially, is formulated in full as Algorithm~\ref{alg:StaticAugmentationScheme}.

\begin{figure}
  \begin{algorithm}[H]
    \caption{Static Channel Augmentation Scheme $\psi_{\rm static}$}
    \label{alg:StaticAugmentationScheme}
    \SetAlgoLined
    \SetKwInOut{Input}{Input}
    \SetKwInOut{Output}{Output}
    \SetKwFor{RepTimes}{repeat}{times}{end}
    \SetKwProg{AugmentationScheme}{Augmentation Scheme}{}{}
    \Input{data set $\mySet{Q}$; \newline augmentation factor $\kappa$} 
    \AugmentationScheme{$(\mySet{Q},\kappa$}{
        $\myVec{\mu}(\mypoint) \gets$ calculate by  \eqref{eq:mean_calculation}, $\forall \mypoint \in \mySet{S}$\;\label{line:centers_calculation}
        
        $\myMat{\Sigma}(\mypoint) \gets$ calculate by  \eqref{eq:variance_calculation}, $\forall \mypoint \in \mySet{S}$\;\label{line:variance_calculation}
        
        $\mydiffSet \gets$ calculate by  \eqref{eq:difference_calculation}, $\forall (\myVec{y}_i^{\rm pilot},\myVec{s}_i^{\rm pilot}) \in \mySet{Q}$\;\label{line:diff_set_calculation}

    Initialize $\mySet{Q}^* = \mySet{Q}$\;
    \RepTimes{$\kappa$}{
    \For{$(\myVec{y}_i,\myVec{s}_i)$ in $\mySet{Q}$}{%
                
                \nonl\texttt{Geometric Augmentation}\;
                sample $\myVec{y}^{\prime}_i$ from $\mathcal{N}(\myVec{\mu}(\mypoint_i),\,\myMat{\Sigma}(\mypoint_i))$\;\label{line:sample_from_distribution}
                add $(\myVec{y}^{\prime}_i,\mypoint_i)$ to buffer $\mySet{Q}^*$; \label{line:add_to_buffer1}

                \nonl\texttt{Constellation-Conserving Projection}\;
                choose a random $\myCCP$ and its corresponding $\myCCPN$\;
                $\myVec{y}^{\prime\prime}_i,\myVec{s}^{\prime\prime}_i \gets$ projection from $\myVec{y}_i, \mypoint_i$ using $\myCCP$ and $\myCCPN$ by \eqref{eq:constellation_conserving_rotation}\;
                add $(\myVec{y}^{\prime\prime}_i,\mypoint^{\prime\prime}_i)$ to buffer $\mySet{Q}^*$; \label{line:add_to_buffer3}
                
                \nonl\texttt{Translation Augmentation}\;
                choose a $\myPTT$\ and a random $\mypoint_i^{\prime\prime\prime} \neq \mypoint_i$\;
                $\Delta \gets \myVec{\mu}(\mypoint_i^{\prime\prime\prime}) - \myPTT\myVec{\mu}(\mypoint_i)$\;\label{line:delta_calculation}
                 $\myVec{y}^{\prime\prime\prime}_i \gets$ calculate from $\myVec{y}_i$, $\Delta$ and $\myPTT$ by  \eqref{eqn:translation2}\;\label{line:difference_addition}
                 add $(\myVec{y}^{\prime\prime\prime}_i,\mypoint_i^{\prime\prime\prime})$ to buffer $\mySet{Q}^*$; \label{line:add_to_buffer2}

                }}
        \KwRet{$\mySet{Q}^*$}
  }
  \end{algorithm}
\end{figure}

\vspace{-0.4cm} 
\subsection{Discussion}
\label{subsec:discussion}

Algorithm~\ref{alg:StaticAugmentationScheme} integrates data augmentation into digital receiver processing, synthesizing channel outputs data from knowledge of the constellation and the available pilots. This augmented data set allows the receiver to train without relying on lengthy pilot transmissions, achieving accurate training with short pilots as empirically demonstrated in Section~\ref{sec:numerical_evaluations}. The suggested  policy differs from augmentations in other domains, e.g., vision, being specifically tailored to communications. The domain knowledge leveraged in our design includes the geometric interpretation of the channel outputs which follows from the typical arrangements of digital constellations; the symmetry of common digital constellations; and the expected symbol diversity, as digital receivers are likely to observe some minimal amount of all possible communication symbols. 

 Contrary to other domains, where augmentations often rely on complex policies, e.g., using generative networks \cite{almahairi2018augmented} or reinforcement learning \cite{laskin2020reinforcement}, we keep our mechanism light. The operations required for our scheme include sampling from a Gaussian distribution, adding and subtracting vectors, and  multiplying by a scalar phase. These operations allow our mechanism to be carried out in real time. In the following section we show how one can also incorporate knowledge regarding the continuous operation of digital receivers, which allows to benefit from data corresponding to  past channels, while still restricting the framework to simple computations.

Algorithm~\ref{alg:StaticAugmentationScheme} can be combined with other approaches for training deep receivers with scarce data. These approaches include the usage of model-based architectures that support the usage of compact \acp{dnn} \cite{shlezinger2020model,farsad2020data}, as well as self-supervision methods \cite{finish2022symbol,shlezinger2019viterbinet}, which assign decision-directed labels to  channel outputs and uses them for training. While self-supervision cannot generate more samples than the block length, Algorithm~\ref{alg:StaticAugmentationScheme} can synthesize any number of samples, and is thus suitable for coping with short block lengths. 
	
	\vspace{-0.2cm}

\section{Data Augmentations for Dynamic Channels}
\label{sec:data_augmentations_dynamic}

\begin{figure}
    \centering
    \includegraphics[width=\columnwidth]{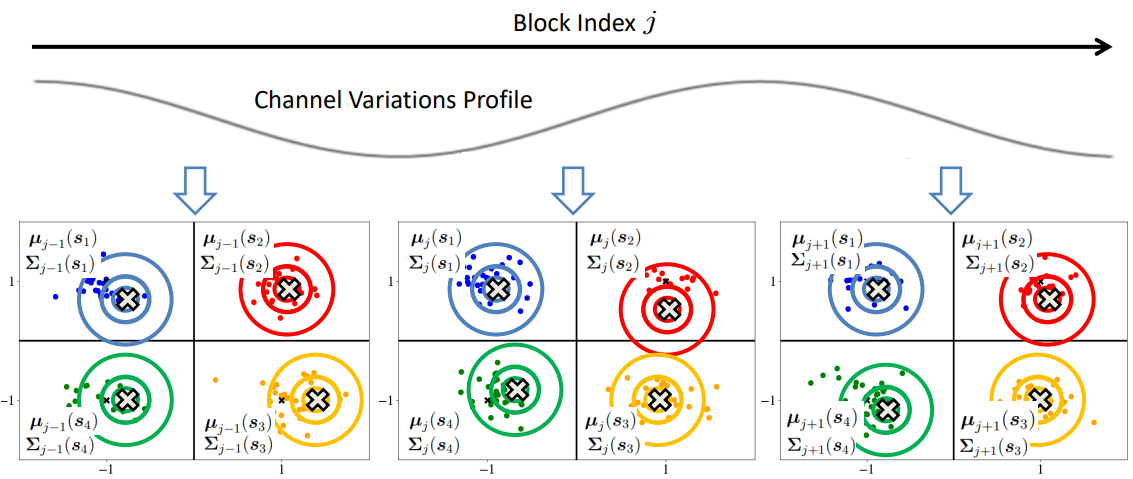}
    \caption{Time-varying channels may benefit from an adaptive augmentation scheme.}
    \label{fig:adaptive_augmentation_illustration}
\end{figure}

\vspace{-0.2cm}
\subsection{Rationale}
\label{subsec:Dyn_Rationale}

Often in practice, digital receivers operate in block-fading channel conditions. In such cases,  each block of $B$ symbols undergoes a different channel. To formulate this, in the sequel we add a subscript $j$ representing the block index, where the block-fading property implies that the statistical relationship between the channel outputs at the $j$th block $\{\myVec{y}_{i,j}\}_{i=1}^{\Blklen}$ and the  symbols $\{\myVec{s}_{i,j}\}_{i=1}^{\Blklen}$ varies with $j$.

While each block requires a different receiver mapping $\Weights_j$, the data augmentation scheme $\psi$ can possibly tune itself between different realizations. For such settings, we are interested in data augmentation schemes that are {\em adaptive}, and are capable of leveraging past channel realizations to improve their ability to synthesize useful training data $\mySet{Q}^*_j$ from the current pilots, as illustrated in Fig.~\ref{fig:adaptive_augmentation_illustration}. Since in this section the proposed scheme is adaptive, the mapping $\psi$ depends not only on the available data $\mySet{Q}_j$, but also on the previously used augmentation, i.e., the mapping applied in the previous block for block fading settings, denoted $\psi_{\rm prev}$.

\vspace{-0.2cm}
\subsection{Adaptive Augmentation}
\label{subsec:Dyn_Augmentation} 
The geometric and translation augmentation techniques proposed in the previous section for static channels both rely on the formulation of clusters, that are used for generating additional synthetic data.  
In block-fading settings, we propose to adaptively smooth the estimated clusters parameters \eqref{eq:mean_calculation} and \eqref{eq:variance_calculation} via a windowed smoothing approach. This way, we combine data across multiple time frames, while reducing the effect of outliers and burst noise during pilot signaling on the estimated clusters, which in turn degrades the usefulness of samples generated from these clusters.

In particular, the motivation stems from the fact that  adjacent blocks may have different number of samples per class in each block, which induces high-variance noise in the estimates of the first and second order statistical moments of the classes. To reduce the noise effects, we exploit the two key properties of digital communications in time-varying channels: 
\begin{enumerate}
    \item \ac{dnn}-aided receivers operate over multiple channel realization, and have thus observed data corresponding to past channel realizations.
\item The variations observed in communication channels are often of a continuous nature.
\end{enumerate} 

Based on the above, we make the augmentation scheme adaptive  by   incorporating  adaptive smoothing. At block index $j$, the clustering mapping utilized by $\psi_{\rm prev}$, denoted $\myVec{\mu}_{j-1}(\cdot)$, and  $\myMat{\Sigma}_{j-1}(\cdot)$, is accounted for when updating the current clustering via
\begin{equation} \label{eq:smoothed_mean_calculation}
    \myVec{\mu}_j(\mypoint) = \frac{\alpha_1}{|\mySet{I}_{\mySet{Q}_j}(\mypoint)|} \sum_{\myVec{y}_{i,j} \in \mySet{Q}_j: i \in  \mySet{I}_{\mySet{Q}_j}(\mypoint)} \myVec{y}_{i,j} + (1-\alpha_1) \cdot \myVec{\mu}_{j-1}(\mypoint) ,
\end{equation}
and
\begin{equation} \label{eq:smoothed_variance_calculation}
    \!\!\myMat{\Sigma}_j(\mypoint) \!=\!\frac{\alpha_2}{|\mySet{I}_{\mySet{Q}_j}(\mypoint)|} \sum_{\myVec{y}_{i,j} \in \mySet{Q}_j: i \in  \mySet{I}_{\mySet{Q}_j}(\mypoint)} \!\!\!\!(\myVec{y}_{i,j}\! - \!\myVec{\mu}_j(\mypoint))(\myVec{y}_{i,j} \!-\! \myVec{\mu}_j(\mypoint))^H + (1-\alpha_2) \cdot \myMat{\Sigma}_{j-1}(\mypoint).
\end{equation}
Here,  $\alpha_1,\alpha_2 \in [0,1]$ are hyperparameters  balancing the contribution of $\psi_{\rm prev}$ on $\psi$. For $\alpha_1=\alpha_2=1$, $\psi$ is not adaptive and is determined solely by $\mySet{Q}_j$, while for $\alpha_1=\alpha_2 = 0$, we have $\psi \equiv \psi_{\rm prev}$.

\begin{figure}
  \begin{algorithm}[H]
    \caption{Dynamic Channel Augmentation Scheme $\psi_{\rm static}$ at block $j$}
    \label{alg:DynamicAugmentationScheme}
    \SetAlgoLined
    \SetKwInOut{Input}{Input}
    \SetKwInOut{Output}{Output}
    \SetKwFor{RepTimes}{repeat}{times}{end}
    \SetKwProg{AugmentationScheme}{Augmentation Scheme}{}{}
    \Input{data set from current block $\mySet{Q}_j$; previous augmentation $\psi_{\rm prev}$ \newline augmentation factor $\kappa$; Smoothing hyperparameters $\alpha_1,\alpha_2$} 
    \AugmentationScheme{$(\mySet{Q}_j,\psi_{\rm prev}, \kappa, \alpha_1. \alpha_2$)}{
        $\myVec{\mu}_j(\mypoint) \gets$ calculate by  \eqref{eq:smoothed_mean_calculation}, $\forall \mypoint \in \mySet{S}$\;\label{line:centers_calculationA}
        
        $\myMat{\Sigma}_j(\mypoint) \gets$ calculate by  \eqref{eq:smoothed_variance_calculation}, $\forall \mypoint \in \mySet{S}$\;\label{line:variance_calculation2}
        
        \textbf{apply steps 4-18 of Algorithm~\ref{alg:StaticAugmentationScheme}} \;\label{lines:pervious_alg}
        
        \KwRet{$\mySet{Q}^*_j$}
  }
  \end{algorithm}
\end{figure}

The dynamic channel augmentation scheme $\psi_{\rm dynamic}$ thus follows the same guidelines as the static augmentaion design of Algorithm~\ref{alg:StaticAugmentationScheme}. The adaptation to temporal variations is accounted for in the estimation of the clusters by replacing lines \ref{line:centers_calculation} and \ref{line:variance_calculation} of Algorithm~\ref{alg:StaticAugmentationScheme} with the calculations in  \eqref{eq:smoothed_mean_calculation} and \eqref{eq:smoothed_variance_calculation}, respectively. The resulting procedure is summarized as Algorithm~\ref{alg:DynamicAugmentationScheme}.	
	\vspace{-0.2cm}
	\section{Numerical Evaluations}
\label{sec:numerical_evaluations}
\vspace{-0.1cm} 
In this section we numerically evaluate the proposed augmentation schemes $\psi_{\rm static}$ and $\psi_{\rm dynamic}$ in finite-memory \ac{siso} channels and in memoryless multi-user \ac{mimo} setups\footnote{The source code used in our experiments is available at \href{https://github.com/tomerraviv95/data-augmentations-for-receivers}{https://github.com/tomerraviv95/data-augmentations-for-receivers}. 
}. 
We first describe the receivers used in our experimental study, detailing their architectures and training settings in Subsection~\ref{subsec:simulated_receivers}. Then, we describe the compared augmented training methods in Subsection~\ref{subsec:augmentation_methods}. After presenting the setup, we introduce the main simulation results for the static augmentation scheme $\psi_{static}$, evaluated on linear synthetic channels, in Subsection~\ref{subsec:static_simulation_results}. Thereafter, we present the evaluation for  the dynamic augmentation scheme $\psi_{dynamic}$ in
Subsection~\ref{subsec:dynamic_linear_synthetic_simulation_results}, Subsection~\ref{subsec:dynamic_cost_simulation_results} and Subsection~\ref{subsec:dynamic_non_linear_synthetic_simulation_results} on synthetic linear channels, synthetic non-linear channels and channels obeying the COST 2100 model, respectively. Finally, we provide ablation study for the contribution of each augmentation  in Subsection~\ref{subsec:ablation_study}.

\vspace{-0.2cm}
\subsection{Evaluated Receivers}
\label{subsec:simulated_receivers}
The \ac{dnn}-aided receiver algorithms used in our experimental study vary based on the considered settings. For each setting, we evaluate our proposed augmentations using both a black-box \ac{dnn} architecture and a hybrid model-based/data-driven receiver.
\subsubsection{Finite-Memory \ac{siso} Channels}
We evaluate \ac{siso} channels, where the channel outputs are scalar, i.e., $N=1$ with a finite memory of $\Mem$ samples. Here, we compare two  \ac{dnn}-based receivers:
\begin{itemize}
    \item The ViterbiNet equalizer, proposed in \cite{shlezinger2019viterbinet}, which is a \ac{dnn}-based Viterbi detector \cite{viterbi1967error} for finite-memory channels. 
    ViterbiNet implements Viterbi equalization in a learned manner by computing the log likelihood metrics utilized in its internal computations using a  \ac{dnn} that requires no  knowledge of the channel  distributions. This internal \ac{dnn} is implemented using three fully-connected layers of sizes $1\times 100$, $100\times 50$, and $50 \times M^\Mem$, with activation functions set to sigmoid (after first layer), ReLU (after second layer), and softmax output layer. %
    \item A recurrent neural network (RNN)  symbol detector, comprised of a sliding-window \ac{lstm} \cite{tandler2019recurrent} with two hidden layers of $64$ cells and a window size of $1$. The output of the \ac{lstm} enters a fully-connected linear layer of size $64 \times M^\Mem$. 
\end{itemize}

\subsubsection{Memoryless \ac{mimo} Channels}
We evaluate two  \ac{dnn}-based \ac{mimo}  receivers for uplink communications with $K$ users:
\begin{itemize}
    \item The DeepSIC receiver proposed in \cite{shlezinger2019deepSIC}. DeepSIC is derived from iterative \ac{sic} \cite{choi2000iterative}, which is a \ac{mimo}  detection method combining multi-stage interference cancellation with soft decisions. DeepSIC here operates in $5$ iterations, refining an estimate of the conditional \acl{pmf} of each symbol based on the soft estimates of the interference symbols using a dedicated \ac{dnn}. 
     DeepSIC is thus comprised of $5 K$ building block \acp{dnn}, which are implemented using three fully-connected layers:  An $(N + K - 1) \times 60$ first layer, a $60 \times 30$ second layer, and a $30 \times M^K$ third layer, with a sigmoid and a ReLU intermediate activation functions, respectively. 
    \item A black-box \ac{dnn} baseline using three fully-connected layers of sizes $N \times 60$, $60\times 60$, and $60 \times M^K$, with ReLU activation functions in-between and softmax output layer.
\end{itemize}

All \ac{dnn}-aided receivers are trained using the Adam optimizer \cite{kingma2014adam} with $I_{\rm sgd}=500$ iterations. The learning rate for ViterbiNet and DeepSIC is set to $10^{-3}$, while for the LSTM black-box and the linear fully-connected black-box it is set to $10^{-2}$. Moreover, the batch size for the LSTM and the fully-connected are set to $16$ and $32$, respectively. These values were set empirically such that the receivers' parameters approximately converge at each time step.  The complete simulation parameters can also be found in the source code that is publicly available online.

\subsection{Augmentation Methods}
\label{subsec:augmentation_methods}
As our focus is on the data used for training \ac{dnn}-aided receivers, we consider the following data sets when training the aforementioned architectures:
\begin{itemize}
    \item {\em Regular training}: No augmentations are applied; each receiver is trained using the original pilots batch $\mySet{Q}$ only.
    \item {\em Combined Scheme}: Our proposed augmentation technique $\psi_{\rm static}$, which includes the sequential application of the geometric, constellation-conserving projection and translation as specified in Algorithm~\ref{alg:StaticAugmentationScheme}. In the case of block-fading channels, as in Subsections~\ref{subsec:dynamic_linear_synthetic_simulation_results}, \ref{subsec:dynamic_cost_simulation_results} and \ref{subsec:dynamic_non_linear_synthetic_simulation_results}, we apply $\psi_{\rm dynamic}$ (Algorithm~\ref{alg:DynamicAugmentationScheme}) with $\alpha_1=\alpha_2=0.3$. 
    \item {\em Extended pilot training}: Here, no augmentations are used, however,  $\beta \times \Blklen^{\rm pilot}$ pilots are transmitted, with $\beta > 1$. Thus, the receiver has more data that is not self-generated for training compared with {\em Regular training}.
\end{itemize}

\vspace{-0.2cm}
\subsection{ Static Linear Synthetic Channels}
\label{subsec:static_simulation_results}
\vspace{-0.1cm}
We begin by evaluating Algorithm~\ref{alg:StaticAugmentationScheme} on stationary synthetic linear channels with additive Gaussian noise, considering  a finite-memory \ac{siso} setting and a memoryless \ac{mimo} setup. As these are relatively simple settings for \ac{dnn}-aided receivers, the purpose of this study is to illustrate the feasibility of the proposed augmentation and its capability of generating useful data, whose gains are further unveiled in the subsequent sections. 
\subsubsection{\ac{siso} Finite-Memory Channels} 
We transmit 100 blocks, each one composed of $\Blklen^{\rm pilot}=200$ pilots followed by $\Blklen^{\rm info}=10,000$ data bits, for a total transmission of a single block with size $\Blklen^{\rm tran}=10,200$. This means that only a small number of pilots are available to train the \ac{dnn}-aided receiver. We consider a real-valued scalar linear Gaussian channel, whose input-output relationship is given by
\begin{equation}
\label{eqn:GaussianSISO}
{y}_i = \sum_{l=0}^{\Mem-1} {h}_{l}{s}_{i-l} + {w}_{i} = \myVec{h}^T \myVec{s}_i + w_i,
\end{equation}
where $\myVec{s}_i$ is the stacking of the last $\Mem$ channel inputs, as defined in Subsection~\ref{subsec:communication_model}. In \eqref{eqn:GaussianSISO}, $\myVec{h} = [1,0.606,0.367,0.223]^T$ are the real channel taps, with a channel memory of $\Mem = 4$ and ${w}_{i}$ is \acl{awgn} with variance $\sigma^2$. The extended pilots method is chosen with $\beta=2.5$, resulting in $500$ total pilots, while the augmentation factor is chosen as $\kappa=3$ for all experiments, resulting in $|\mySet{Q}^*| = 10 \cdot |\mySet{Q}|$.

\begin{figure*}
    \centering
    \begin{subfigure}[b]{0.48\textwidth}
    \includegraphics[width=\textwidth,height=0.24\textheight]{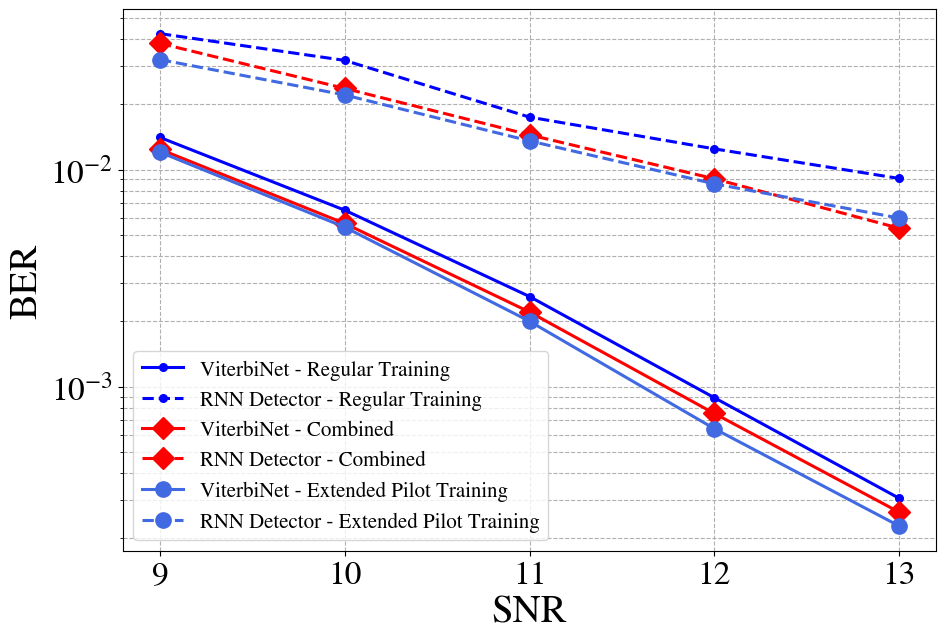}
    \caption{\ac{siso} - \ac{ber} vs. SNR.}
    \label{fig:SNR_linear_SISO_trial_1}
    \end{subfigure}
    \begin{subfigure}[b]{0.48\textwidth}
    \includegraphics[width=\textwidth,height=0.24\textheight]{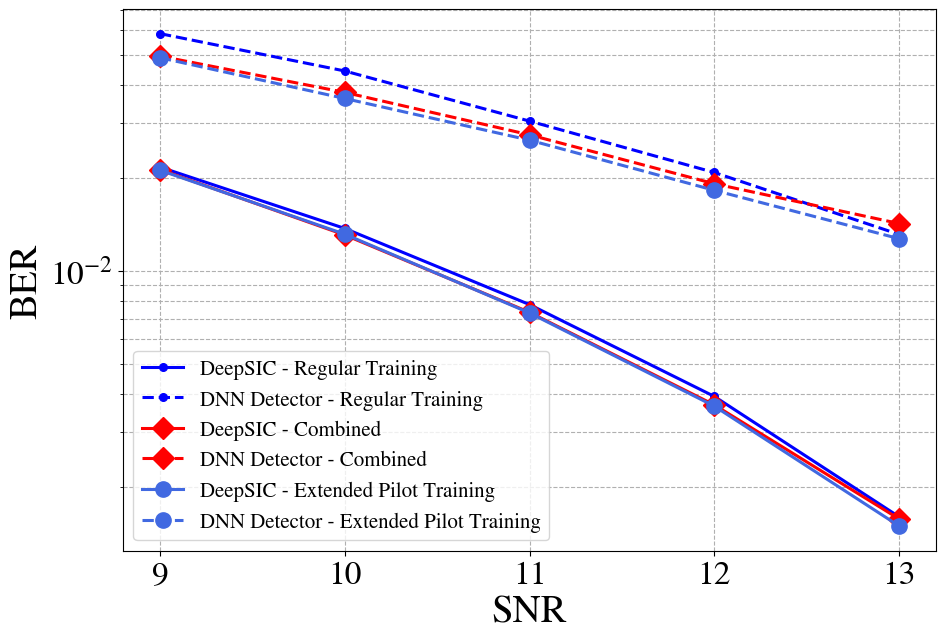}
    \caption{\ac{mimo} - \ac{ber} vs. SNR.}
    \label{fig:SNR_linear_MIMO_trial_1}
    \end{subfigure}
    \caption{Results on static synthetic linear Gaussian channel.}
    \label{fig:static_linear_synthetic_BER_vs_SNR} 
    \figSpace
\end{figure*}

In Fig.~\ref{fig:SNR_linear_SISO_trial_1} we plot the \ac{ber} evaluated via \eqref{eqn:ErrorRate}, averaged over the transmission of the 100 mentioned blocks, when the \ac{snr}, defined as $1/\sigma^2$, takes values in the range of $9$ dB - $13$ dB. This figure shows that the combined augmentation approach $\psi_{\rm static}$ outperforms the regular training,  demonstrating gains of around $0.1$ dB for the ViterbiNet (which is known to be amenable to adaptation with small training sets \cite{shlezinger2019viterbinet}), and up to $1$ dB in the black-box RNN case. Furthermore, data augmentation allows to approach the performance achieved with extended pilots, i.e., with $2.5\times$ more pilots. 

\subsubsection{Memoryless MIMO Channels} 
We next consider a memoryless \ac{mimo} setting. Here, we set the number of symbols in each block to $\Blklen^{\rm pilot} = 600$, $\Blklen^{\rm info} = 10,000$ and $\Blklen^{\rm tran} = 10,600$. 
The input-output relationship of the memoryless Gaussian \ac{mimo} channel is given by
\begin{equation}
\label{eqn:GaussianMIMO}
\myVec{y}_i = \myMat{H}\myVec{s}_i + \myVec{w}_i,
\end{equation}
where $\myMat{H}$ is a known deterministic $N\times K$ channel matrix, and $\myVec{w}_i$ is complex Gaussian noise with covariance $\sigma^2\myMat{I}_N$.  
We set the number of users and antennas to $K=N=4$. The channel matrix $\myMat{H}$ models spatial exponential decay, and its entries are given by
$\left( \myMat{H}\right)_{n,k} = e^{-|n-k|}$, for each $n \in \{1,\ldots, N\}$, $ k \in \{1,\ldots, K\}$.  The transmitted symbols are generated from a \ac{qpsk} constellation in a uniform i.i.d. manner, i.e., $\mathcal{S}=\{(\pm \frac{1}{\sqrt{2}},\pm \frac{1}{\sqrt{2}})\}^4$. Here, the extended pilot method is simulated with $\beta = 1.75$, resulting in $1,050$ pilots in total, while the augmentation factor is chosen as $\kappa=2$ for all experiments, resulting in $|\mySet{Q}^*| = 7 \cdot |\mySet{Q}|$.

As observed in Fig.~\ref{fig:SNR_linear_MIMO_trial_1},  in this synthetic setting the usage of data augmentations allows one to  achieve performance similar to that of increasing the number of pilots by a factor of $1.75$. In particular, since the hybrid model-based/data-driven DeepSIC only needs short pilot blocks to reliably estimate the networks' parameters under the simple synthetic linear Gaussian case \cite{shlezinger2019deepSIC}, the addition of either augmentations or more pilots is negligible (up to $0.05$ dB). For the black-box architecture which is more data-hungry, the augmentations yield gains of up to $0.5$ dB, reaching close to the extended pilots baseline, in the low-to-medium SNR regime. 

The above figures serve to show that our augmentations are tailored to the detection task, being agnostic to the actual structure of the receiver, and offer small but consistent improvements even in the trivial case of static linear channels.

\begin{figure*}
    \centering
    \begin{subfigure}[b]{0.48\textwidth}
    \includegraphics[width=\textwidth,height=0.2\textheight]{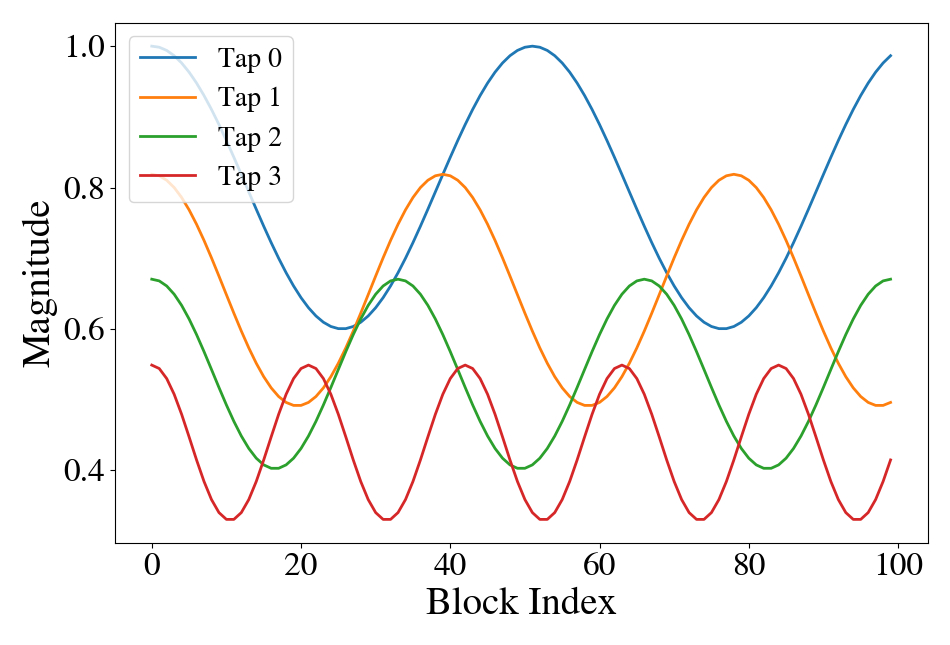}
    \caption{Synthetic channel.}
    \label{fig:synthetic_channel_taps_siso}
    \end{subfigure}
    \begin{subfigure}[b]{0.48\textwidth}
    \includegraphics[width=\textwidth,height=0.2\textheight]{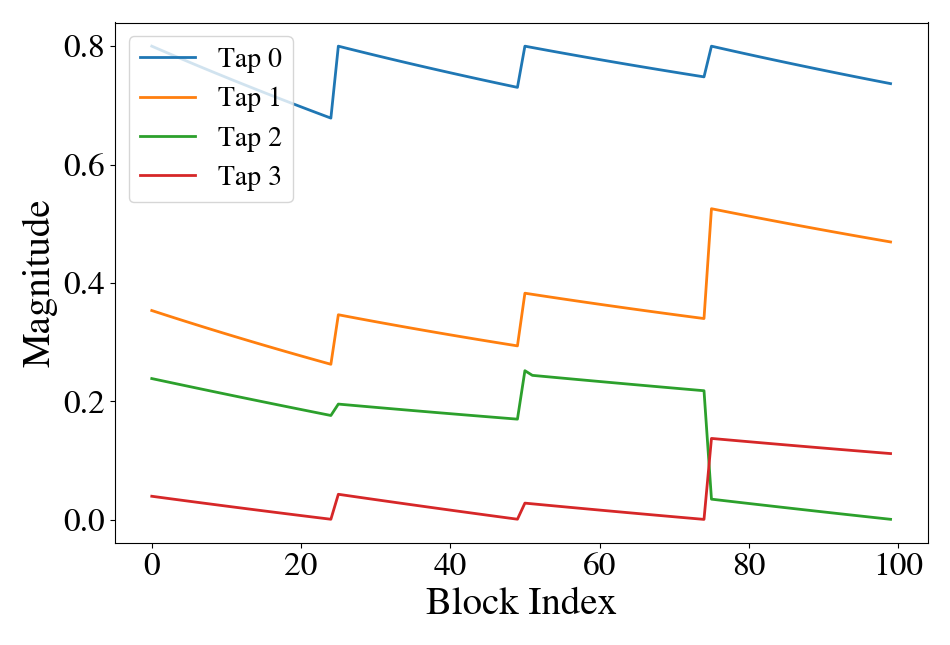}
    \caption{COST 2100 channel.}
    \label{fig:cost_channel_taps_siso} 
    \end{subfigure}
    \caption{Time-varying \ac{siso} channels examples: channel coefficients versus block index.}
    \figSpace
    \label{fig:channels} 
\end{figure*}

\vspace{-0.2cm}
\subsection{Time-Varying Linear Synthetic Channels}
\label{subsec:dynamic_linear_synthetic_simulation_results}
\vspace{-0.1cm}

We proceed by evaluating Algorithm~\ref{alg:DynamicAugmentationScheme} in block-fading synthetic linear channels with additive Gaussian noise. We again consider  both the \ac{siso} or the \ac{mimo} setups. 
We are further interested in measuring the effective gains of our as a function of the transmitted pilots. Obviously, as the number of pilots transmitted increases, the gain of additional synthetic examples diminishes. Thus, in the following study we also quantify the effective gain of our approach.
\subsubsection{\ac{siso} Finite-Memory Channels} 
We simulate a block-fading channel using the same settings of $\Blklen^{\rm pilot}, \Blklen^{\rm info}$ and $\Blklen^{\rm tran}$ as used in Subsection~\ref{subsec:static_simulation_results}. For the extended pilots and combined augmentations methods, we set $\beta=2.5$ and $\kappa=3$. The input-output relationship is the same as in \eqref{eqn:GaussianSISO}. Yet, here the channel taps $\myVec{h}$ vary between blocks following the profile depicted in Fig.~\ref{fig:synthetic_channel_taps_siso}, which is a synthetic model representing oscillations of varying frequencies. 

In Fig.~\ref{fig:SNR_linear_synth_SISO_fading_trial_1} we plot the average \ac{ber}, as in  \eqref{eqn:ErrorRate}, for the different considered settings of $\mySet{Q}^*$, when the \ac{snr}  varies in the range of $9$ dB - $13$ dB. We observe in  Fig.~\ref{fig:SNR_linear_synth_SISO_fading_trial_1} that in the block-fading linear case, the combined approach allows the \ac{dnn}-aided receivers to approach the performance achieved when training with extended pilots over all SNRs. In high SNRs, the gains increase, being significant for both the black-box RNN detector and the model-based ViterbiNet.

We next evaluate the gains in pilot efficiency, employing the proposed augmentation with $\kappa=3$. Here, we compare the \ac{ber} of the considered \ac{dnn}-aided receivers for different number of pilots $|\mySet{Q}|$ for fixed \ac{snr} of $12$ dB. The results, presented in Fig.~\ref{fig:pilots_size_SISO_study},  show that the proposed augmentations allow both \ac{dnn}-aided receivers to improve their pilot efficiency by factors of te order of $2\times$ and $3\times$. For instance, the RNN detector requires at least $600$ pilots to achieve a similar \ac{ber} to that which it achieves when augmenting merely $200$ pilots using the proposed method. 


\subsubsection{Memoryless MIMO Channels} 
For the \ac{mimo} setting, we use the same values of $\Blklen^{\rm pilot}$, $\Blklen^{\rm info}$,$\Blklen^{\rm tran}$,$\beta$ and $\kappa$, as set in Subsection~\ref{subsec:static_simulation_results}. 
The input-output relationship of the channel follows  \eqref{eqn:GaussianMIMO} with $K=N=4$ and symbols drawn from a \ac{qpsk} constellation. Here, each channel from the $K=4$ users to one of the $N=4$ antennas undergoes the blockwise-variations  profile illustrated in Fig.~\ref{fig:synthetic_channel_taps_siso}. This models a loss of information between each set of transmissions and the corresponding receiver in an i.i.d. manner.

\begin{figure*}
    \centering
    \begin{subfigure}[b]{0.48\textwidth}
    \includegraphics[width=\textwidth,height=0.24\textheight]{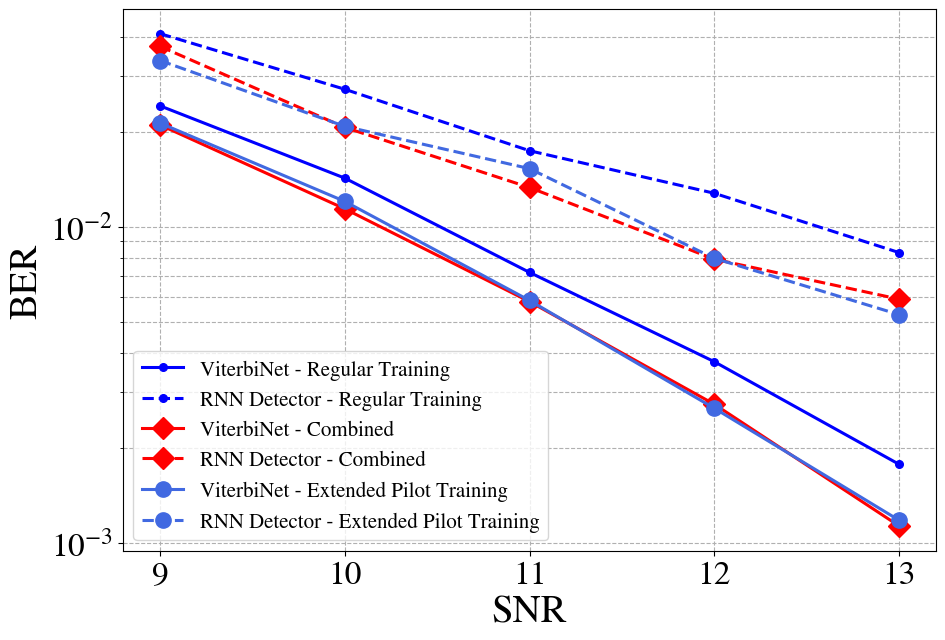}
    \caption{\ac{siso} - \ac{ber} vs. SNR.}
    \label{fig:SNR_linear_synth_SISO_fading_trial_1}
    \end{subfigure}
    \begin{subfigure}[b]{0.48\textwidth}
    \includegraphics[width=\textwidth,height=0.24\textheight]{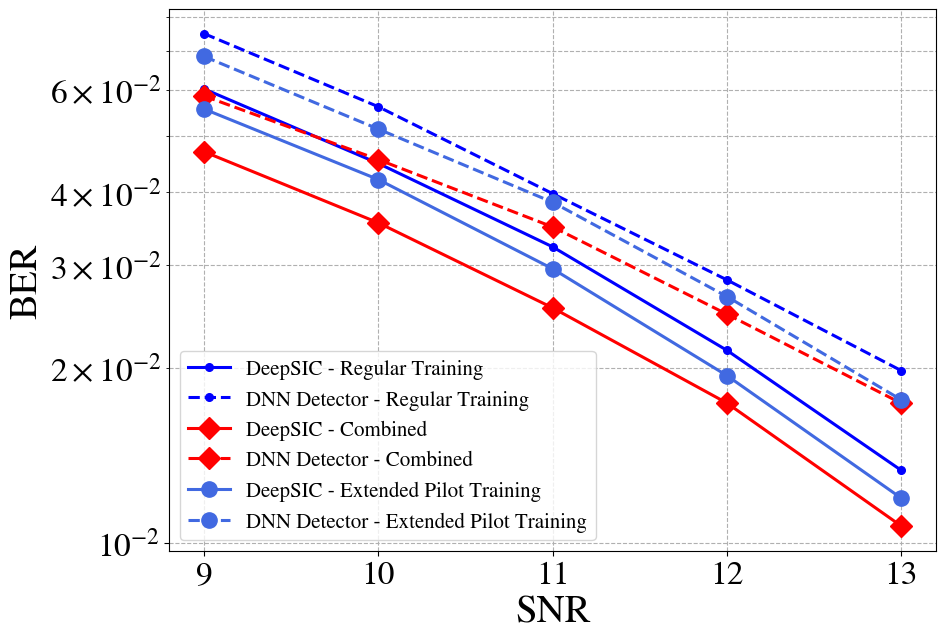}
    \caption{\ac{mimo} - \ac{ber} vs. SNR.}
    \label{fig:SNR_linear_synth_MIMO_fading_trial_1}
    \end{subfigure}
    \caption{Results on time-varying synthetic linear Gaussian channel.}
    \label{fig:time_varying_linear_synthetic_BER_vs_SNR} 
    \figSpace
\end{figure*}

As we observe in Fig.~\ref{fig:SNR_linear_synth_MIMO_fading_trial_1}, the gains of enriching a data set via the proposed augmentations here are larger than the static case. This stems from the fact that  the addition of either pilots/augmentations aids in matching the momentary training distribution with the changing blockwise changes in the channel response. Both DeepSIC and the black-box DNN detector benefit from a gain of around $0.5$ dB by training with our proposed augmentations scheme, over the regular training methods. The gains remain relatively consistent over all evaluated SNRs. One important highlight is that our method is able to surpass even the case of sending more pilots, as depicted by the extended pilots training method, and can thus simultaneously improve both spectral efficiency as well as \ac{ber} performance.

\begin{figure*}
    \centering
    \begin{subfigure}[b]{0.48\textwidth}
    \includegraphics[width=\textwidth,height=0.24\textheight]{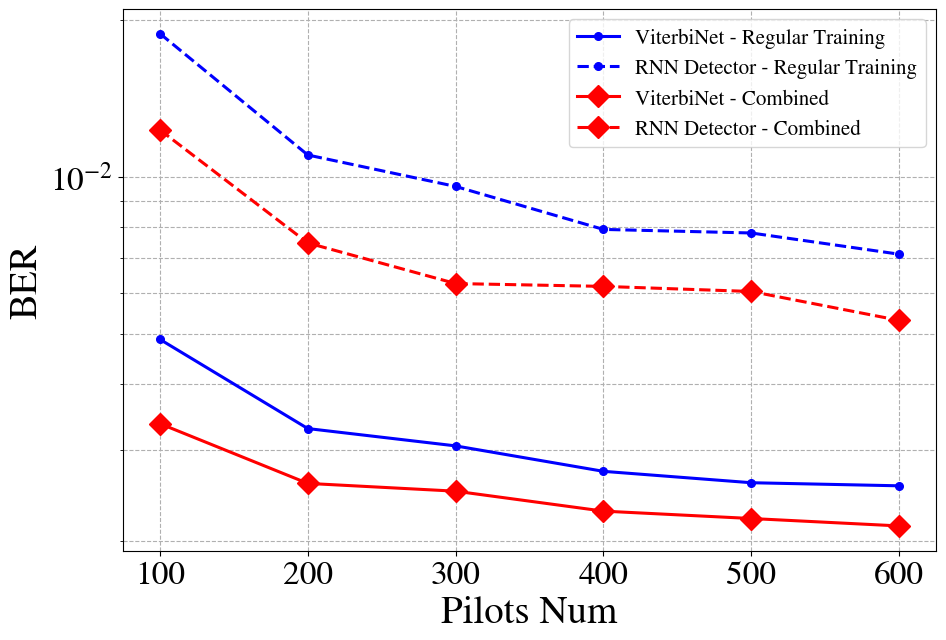}
    \caption{\ac{siso} - \ac{ber} vs. transmitted pilots.}
    \label{fig:pilots_size_SISO_study}
    \end{subfigure}
    \begin{subfigure}[b]{0.48\textwidth}
    \includegraphics[width=\textwidth,height=0.24\textheight]{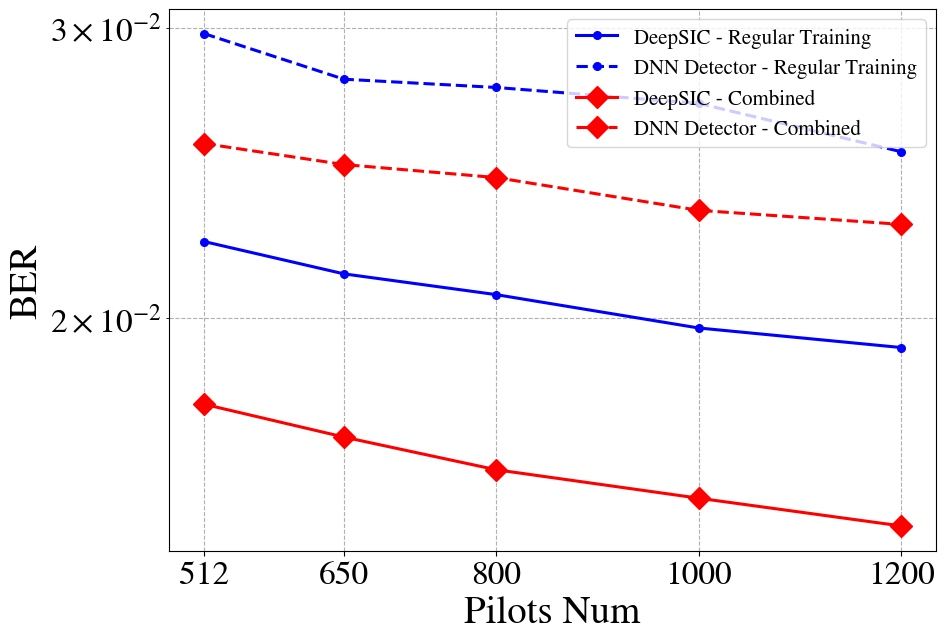}
    \caption{\ac{mimo} - \ac{ber} vs. transmitted pilots.}
    \label{fig:pilots_size_MIMO_study}
    \end{subfigure}
    \caption{Pilots efficiency study on time-varying synthetic linear Gaussian channel}
    \label{fig:pilots_size_study} 
    \figSpace
\end{figure*}

To verify the contribution of the different augmentations on the more complex \ac{mimo} setup, we evaluate its improvements in pilot efficiency when generating data while setting $\kappa= 2$. The resulting \ac{ber} versus number of pilots $|\mySet{Q}|$ for \ac{snr} of $12$ dB are reported in Fig.~\ref{fig:pilots_size_MIMO_study}. We observe here notable gains in pilot efficiency for both \ac{dnn}-aided receivers. In particular, the \ac{ber} achieved with data augmentation surpasses that of the regular training method by a factor of around $2 \times$ (similar to the above \ac{siso} case) for both the model-based deep receiver and the black-box architecture. This spectral efficiency factor remains almost constant even as one increases the pilots length significantly, with no foreseeable saturation.

\begin{figure*}
    \centering
    \begin{subfigure}[b]{0.48\textwidth}
    \includegraphics[width=\textwidth,height=0.24\textheight]{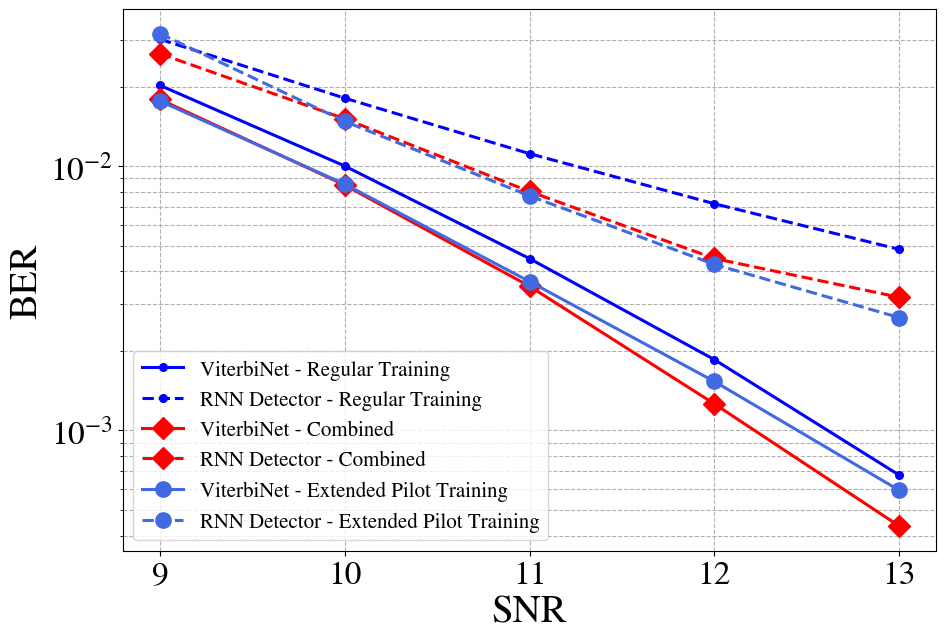}
    \caption{\ac{siso} - \ac{ber} vs. SNR.}
    \label{fig:SNR_linear_COST_2100_SISO}
    \end{subfigure}
    \begin{subfigure}[b]{0.48\textwidth}
    \includegraphics[width=\textwidth,height=0.24\textheight]{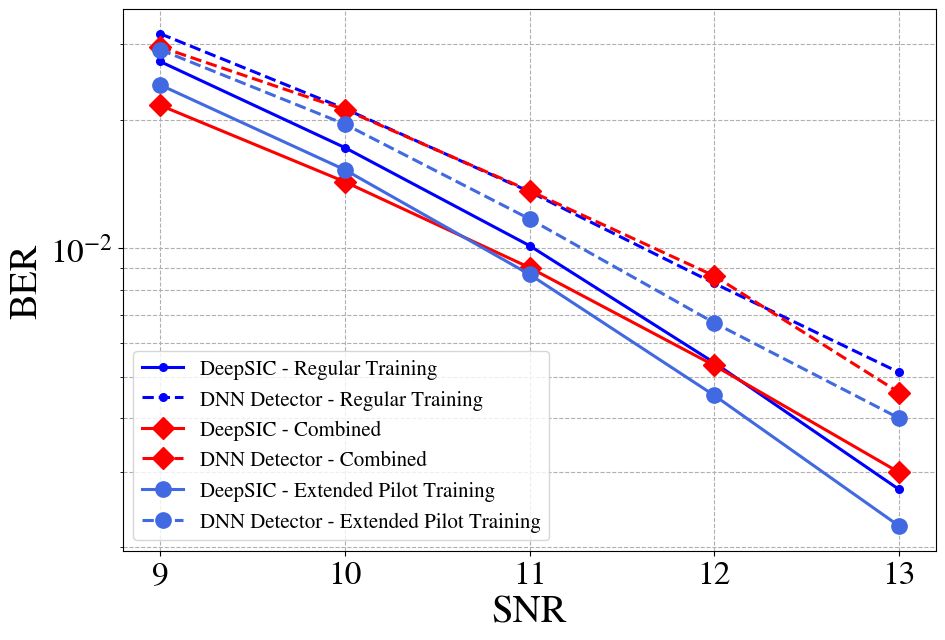}
    \caption{\ac{mimo} - \ac{ber} vs. SNR.}
    \label{fig:SNR_linear_COST_2100_MIMO}
    \end{subfigure}
    \caption{Results on time-varying COST 2100 linear Gaussian channel}
    \label{fig:time_varying_linear_cost_2100_BER_vs_SNR} 
    \figSpace
\end{figure*}

\vspace{-0.2cm}
\subsection{Time-Varying Linear COST Channels}
\label{subsec:dynamic_cost_simulation_results}
\vspace{-0.1cm}
Next, we evaluate the proposed augmentation scheme in time-varying linear channels, where we employ the COST 2100 model for generating the time-varying taps. The COST 2100 geometry-based stochastic channel model \cite{liu2012cost} is a widely-used model to represent wireless communications scenarios.

\subsubsection{\ac{siso} Finite-Memory Channels} We employ a similar linear model as in \ref{eqn:GaussianSISO}, but incorporate a time varying characteristics: 
\begin{equation}
\label{eqn:GaussianSISOvarying}
{y}_{i,j} = \sum_{l=0}^{\Mem-1} {h}_{l,j}{s}_{i-l,j} + {w}_{i,j} = \myVec{h}_{j}\myVec{s}_{i,j} + {w}_{i,j}.
\end{equation}
The taps realizations $\myVec{h}_j = [{h}_{0,j},\ldots,{h}_{\Mem-1,j}]^T$ are generated using the wideband indoor hall $5$ GHz setting of COST2100 with single-antenna elements, and ${w}_{i,j}$ is  an \acl{awgn} with variance $\sigma^2$. The resulting channel taps are displayed in Fig.~\ref{fig:cost_channel_taps_siso}. This setting represents a user moving in an indoor setup while switching between different microcells. Succeeding on this scenario requires high adaptivity since there is considerable mismatch in the channels observed between different blocks. 

We transmit 100 blocks, and measure the average \ac{ber} over all blocks and bits. The same configuration as before holds; with values of $\Blklen^{\rm pilot} = 200$, $\Blklen^{\rm info} = 10,000$ (that is $\Blklen^{\rm tran} = 10,200$), $\beta=2.5$, and $\kappa=3$. Fig.~\ref{fig:SNR_linear_COST_2100_SISO} shows the average results over these 100 blocks. We observe that gains of up to $1$ dB are achieved by employing our scheme, converging  with the performance of the extended pilots training for the black-box architecture, while surpassing this baseline for the hybrid model-based/data-driven ViterbiNet. These gains incur only small computational cost. Specifically, they are achieved at a low cost, relying on  only the simulation of additional data for training, while the same training complexity, i.e., number of iterations, is applied under each training method.

We proceed to evaluate pilot efficiency on the realistic COST channel to further validate our method. The results for \ac{snr} of $12$ dB are reported in  Fig.~\ref{fig:pilots_size_SISO_cost_study}, which shows that for the \ac{siso} ViterbiNet and RNN cases, the spectral efficiency factor decreases as the pilots number increases, down to a factor of $\times 2.5$ at high pilots number. That is, for $\Blklen^{\rm pilot}=200$, the training of the augmented versions is approximately equal to the performance achieved at $\Blklen^{\rm pilot}=500$ for non-augmented regular training. Moreover, while saturation is achieved quickly with the regular trained ViterbiNet architecture, the augmented training is able to further decrease the error rate with an increase of the pilots, contributing to the overall data diversity one relies on.

\subsubsection{Memoryless \ac{mimo} Channels} Similarly to the \ac{siso} case, we simulate a time-varying linear \ac{mimo} channel from \eqref{eqn:GaussianMIMO} via
\begin{equation}
\label{eqn:GaussianMIMOvarying}
\myVec{y}_{i,j} = \myMat{H}_j\myVec{s}_{i,j} + \myVec{w}_{i,j},
\end{equation}
where the subscript $j$ expresses the block index dependence of each term. The instantaneous channel matrix $\myMat{H}_j$ is simulated by following the above \ac{siso} description, and creating $4 \times 4 = 16$ narrow-band channels with  COST 2100 using the $5$GHz indoor configuration.

As this scenario is harder to track by any online training method, we transmit more pilots than before, setting $\Blklen^{\rm pilot} = 1,000$ , $\Blklen^{\rm info} = 10,000$ (that is $\Blklen^{\rm tran} = 11,000$). All other hyperparameters are as in Subsection~\ref{subsec:static_simulation_results} for the \ac{mimo} case. The \ac{ber} results are calculated after the transmission of 100 blocks, and the resulting \ac{ber} curves are reported in Fig.~\ref{fig:SNR_linear_COST_2100_MIMO}. We observe in Fig.~\ref{fig:SNR_linear_COST_2100_MIMO}  gains of up to $0.6$ dB in the low-to-medium SNR regime in the DeepSIC case, and in the high SNR regime for the black-box DNN a small gain of $0.1$ dB.

\begin{figure*}
    \centering
    \begin{subfigure}[b]{0.48\textwidth}
    \includegraphics[width=\textwidth,height=0.24\textheight]{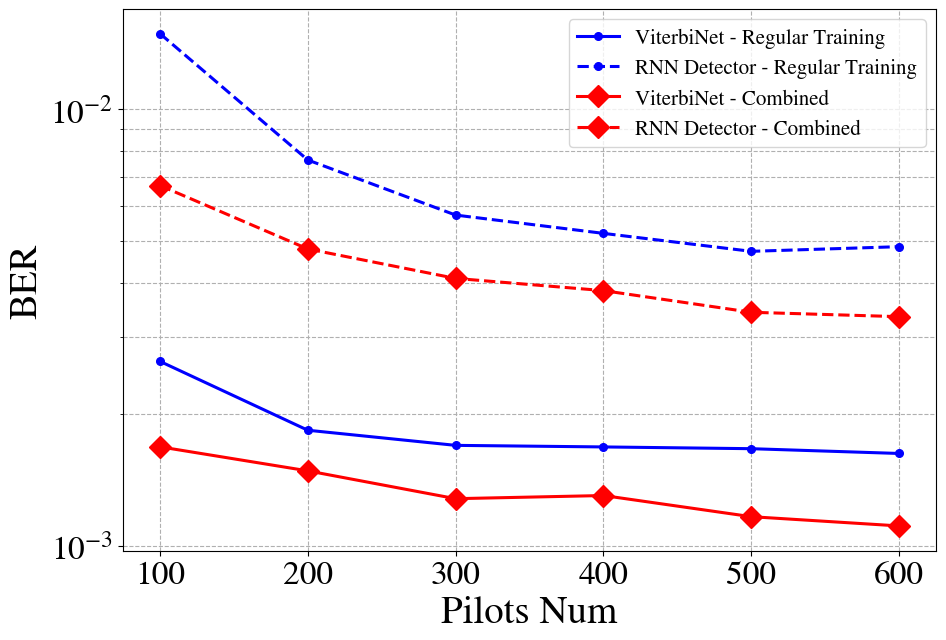}
    \caption{\ac{siso} - \ac{ber} vs. transmitted pilots.}
    \label{fig:pilots_size_SISO_cost_study}
    \end{subfigure}
    \begin{subfigure}[b]{0.48\textwidth}
    \includegraphics[width=\textwidth,height=0.24\textheight]{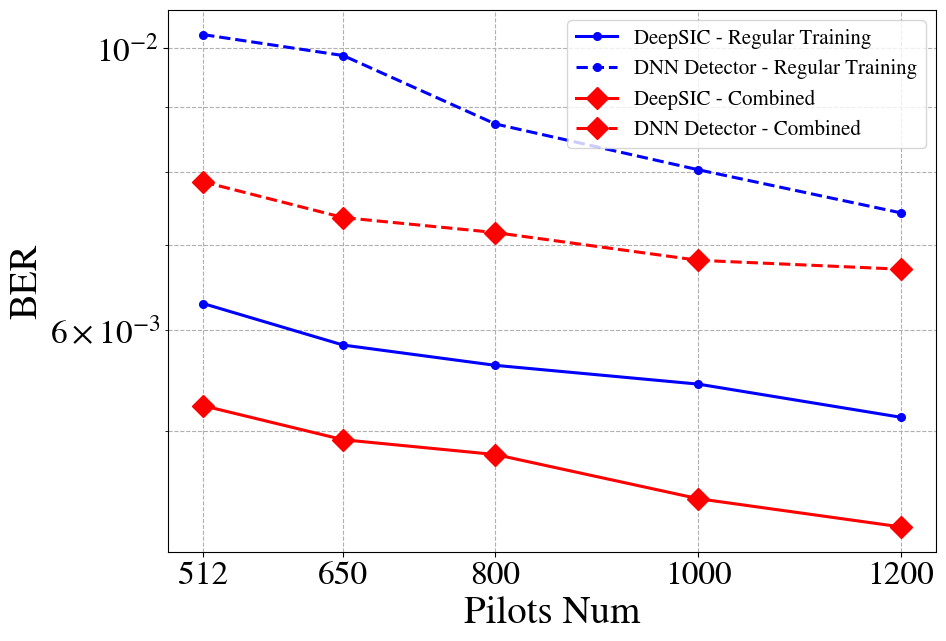}
    \caption{\ac{mimo} - \ac{ber} vs. transmitted pilots.}
    \label{fig:pilots_size_MIMO_cost_study}
    \end{subfigure}
    \caption{Pilots efficiency study on time-varying COST 2100 linear Gaussian channel}
    \label{fig:pilots_size_cost_study} 
    \figSpace
\end{figure*}

To quantify the spectral efficiency gain, in Fig.~\ref{fig:pilots_size_MIMO_cost_study} we observe that the \ac{ber} of DeepSIC or the \ac{dnn} detector trained regularly with $1200$ pilot symbols is similar to the one achieved with augmented training over $650$ pilot symbols only, corresponding to a gain factor of almost $\times 2$. This illustrates that our proposed augmentation is indeed beneficial on common channel models, and not only with simplistic synthetic ones.

\begin{figure*}
    \centering
    \begin{subfigure}[b]{0.48\textwidth}
    \includegraphics[width=\textwidth,height=0.24\textheight]{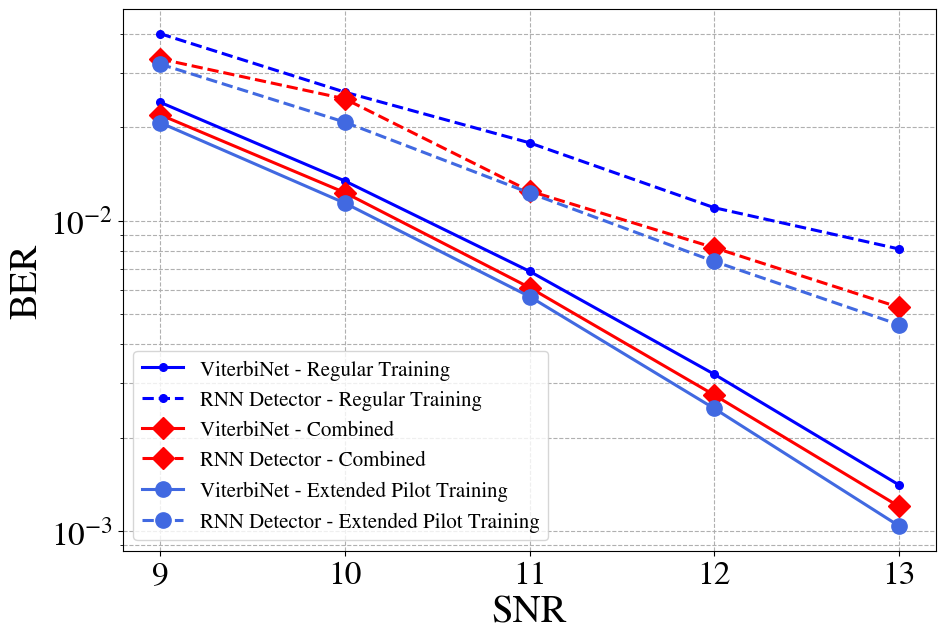}
    \caption{\ac{siso} - \ac{ber} vs. SNR.}
    \label{fig:SNR_non_linear_synth_SISO_fading}
    \end{subfigure}
    \begin{subfigure}[b]{0.48\textwidth}
    \includegraphics[width=\textwidth,height=0.24\textheight]{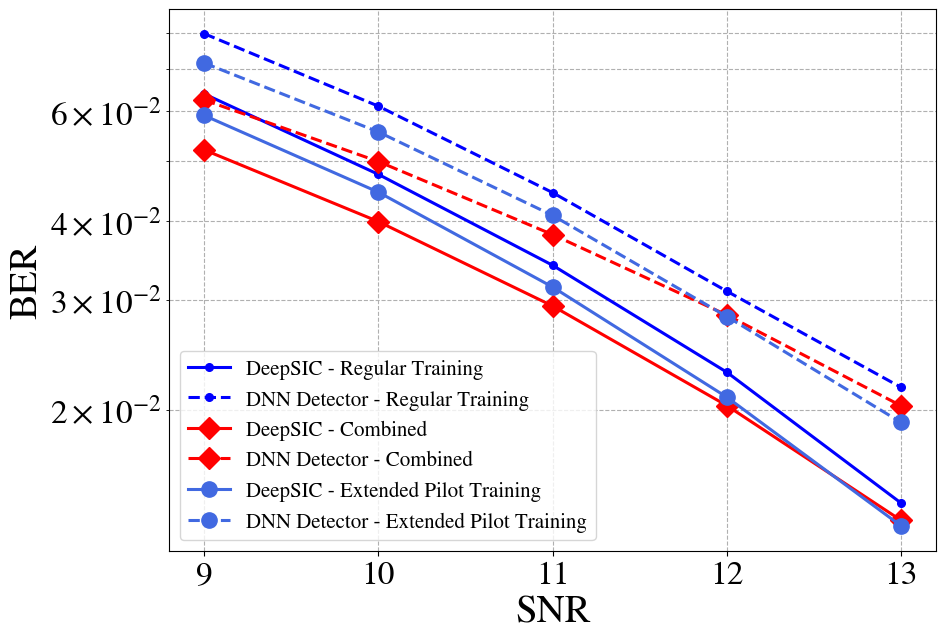}
    \caption{\ac{mimo} - \ac{ber} vs. SNR.}
    \label{fig:SNR_non_linear_synth_MIMO_fading}
    \end{subfigure}
    \caption{Results on time-varying synthetic non-linear Gaussian channel}
    \label{fig:time_varying_non_linear_synthetic_BER_vs_SNR} 
    \figSpace
\end{figure*}

\vspace{-0.2cm}
\subsection{Time-Varying Non-Linear Synthetic Channels}
\label{subsec:dynamic_non_linear_synthetic_simulation_results}
\vspace{-0.1cm}
Further evaluation of the proposed online enrichment scheme is done on non-linear channels. These channels are of particular interest, as one of the motivations for \ac{dnn}-based receivers arises from the need to cope with complex channel models, as classical algorithm usually assume a linear Gaussian channel and perform rather poorly on these non-linear cases.

\subsubsection{\ac{siso} Finite-Memory Channels} To check the robustness of our method in a non-linear case, we generate such a channel by applying a non-linear transformation to the synthetic channel in \eqref{eqn:GaussianSISO}. The resulting finite-memory \ac{siso} channel is given by
\begin{equation}
\label{eqn:Gaussian1}
{y}_{i,j} =\tanh{\Big(C \cdot \Big(\sum_{l=0}^{\Mem-1} h_{l,j}{s}_{i-l,j} + w_{i,j}\Big)\Big)}.
\end{equation}
This operation may represent, e.g., non-linearities induced by the receiver acquisition hardware. The hyperparameter $C$ stands for a power attenuation at the receiver, and was chosen empirically as $C=1$. All other hyperparameters and settings are the same as those used in the previous subsections. The channel taps follow a fading profile as depicted in Fig.~\ref{fig:synthetic_channel_taps_siso}.  

The simulation results depicted in Fig.~\ref{fig:SNR_non_linear_synth_SISO_fading}  empirically show that a consistent gain of up to $0.2$ dB can be achieved by employing our scheme for the ViterbiNet receiver, or $1$ dB for the black-box RNN case. Comparing with the linear case presented in Subsection~\ref{subsec:static_simulation_results}, we see that the overall gains by either extended pilots training, or by our proposed combined augmentations scheme, are comparable and  consistent over  all the \ac{snr} range.

\subsubsection{Memoryless \ac{mimo} Channels} We simulate a non-linear \ac{mimo} channel from \eqref{eqn:GaussianMIMO} via
\begin{equation}
\label{eqn:Gaussian2}
\myVec{y}_{i,j} = \tanh{\Big(C \cdot\Big(\myMat{H}_j\myVec{s}_{i,j} + \myVec{w}_{i,j}\Big)\Big)},
\end{equation}
with $C=1$ and with simulation settings that are chosen as before in Subsection~\ref{subsec:static_simulation_results}. The numerical results for this channel are illustrated in Fig.~\ref{fig:SNR_non_linear_synth_MIMO_fading}, showing that the superiority of our approach is maintained in non-linear setups. Compared to the linear case, we observe higher gains of up to $0.75$ dB of our scheme in low-to-medium SNRs, and up to $0.2$ dB in high SNRs. 

\vspace{-0.2cm}
\subsection{Ablation Study of Different Augmentations}
\label{subsec:ablation_study}
\vspace{-0.1cm}

To explore how each much each augmentation contributes to the final combined approach, we next evaluate the gains achieved by the addition of each  of the proposed augmentation techniques. We refer to the different augmentations proposed for static channels as \textit{Geometric Augmentation}, \textit{Constellation-Conserving Rotation Augmentation} and finally the \textit{Translation Augmentation}. Each one of these augmentation techniques is compared with regular non-augmented training, as well as with the final combined approach, which employs all three augmentations sequentially as specified in Algorithm~\ref{alg:DynamicAugmentationScheme}. We again consider both \ac{siso} and \ac{mimo} settings.

\subsubsection{\ac{siso} Finite-Memory Channels} We employ the time-varying linear synthetic channels setup from Subsection~\ref{subsec:static_simulation_results}, including all hyperparameters. For the single augmentation methods, we set the augmentation factor as $\kappa=9$, resulting in $|\mySet{Q}^*| = 10 \cdot |\mySet{Q}|$ in total. This means that the number of samples in the single augmentation case is equal to the number of augmented samples in the combined case with $\kappa=3$. In Fig.~\ref{fig:ablation_SNR_linear_synth_SISO_fading} we plot the \ac{ber} versus the \ac{snr} value. It is observed in Fig.~\ref{fig:ablation_SNR_linear_synth_SISO_fading}  that the geometric augmentation dominates over the two different ones, dictating the combined \ac{ber}, up to high SNR.

\subsubsection{Memoryless \ac{mimo} Channels} 
\begin{figure*}
    \centering
    \begin{subfigure}[b]{0.48\textwidth}
    \includegraphics[width=\textwidth,height=0.24\textheight]{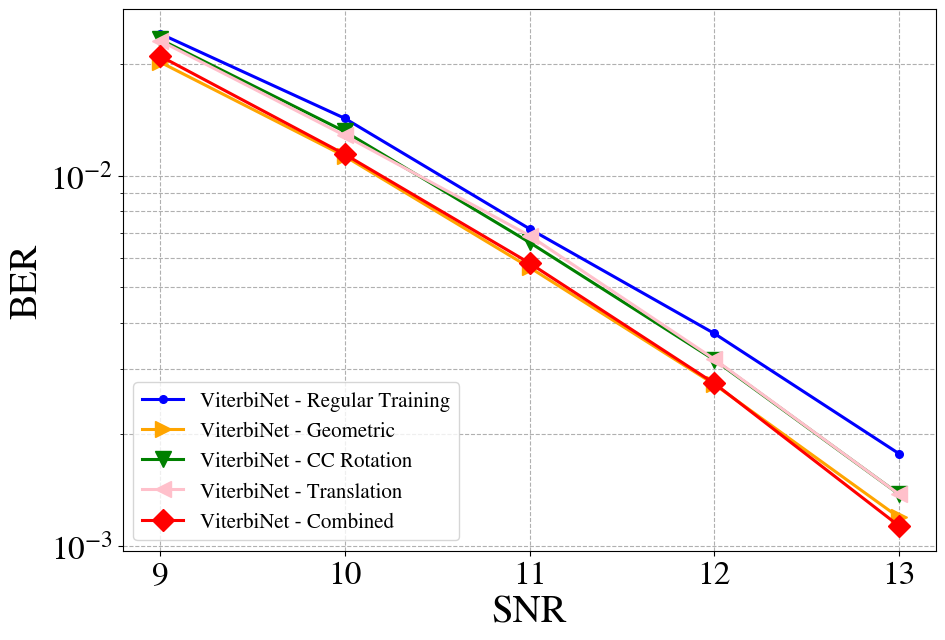}
    \caption{\ac{siso} - \ac{ber} vs. SNR.}
    \label{fig:ablation_SNR_linear_synth_SISO_fading}
    \end{subfigure}
    \begin{subfigure}[b]{0.48\textwidth}
    \includegraphics[width=\textwidth,height=0.24\textheight]{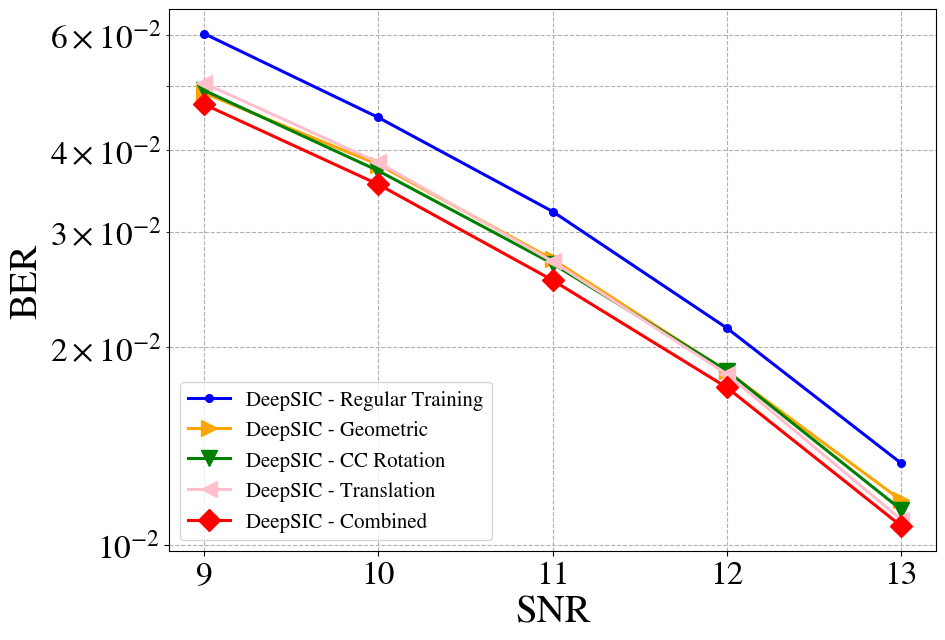}
    \caption{\ac{mimo} - \ac{ber} vs. SNR.}
    \label{fig:ablation_SNR_linear_synth_MIMO_fading}
    \end{subfigure}
    \caption{Ablation study on time-varying synthetic linear Gaussian channel}
    \label{fig:ablation_time_varying_linear_synthetic_BER_vs_SNR} 
    \figSpace
\end{figure*}

To verify the contribution of the different augmentations on the more complex \ac{mimo} setup, we again employ the time-varying linear synthetic channels setup from Subsection~\ref{subsec:static_simulation_results}, including all hyperparameters. The augmentation factor is chosen as $\kappa=6$ for all single augmentation methods, again matching the number of samples in the single augmentation case to the combined augmentations approach. In Fig.~\ref{fig:ablation_SNR_linear_synth_MIMO_fading} we observe that the \ac{ber} of DeepSIC over the different augmentations surpasses that of the regular training method. Each proposed augmentation surpasses the regular training by around $0.5$ dB, with the gains adding up in the combined approach, up to $0.7$ dB gain over the regular training. These results demonstrate that  the proposed augmentations are complementary of another, and that combining them via Algorithm~\ref{alg:StaticAugmentationScheme} is indeed beneficial.

	\vspace{-0.2cm}
	\section{Conclusion}
	\label{sec:conclusion}
	\vspace{-0.1cm}
In this paper, we proposed novel data augmentations techniques tailored specifically to facilitate training of \ac{dnn}-aided receivers with limited data. The novelty originates from the a-priori structure of digital constellations, as well as the dynamic nature of communication channel, which enforce the augmentations to satisfy the diversity and low complexity principles. We numerically evaluated the proposed schemes and showed that significant gains, of up to $1$dB in \ac{ber} and of up to $\times 3$ in spectral efficiency, can be achieved over numerous linear and non-linear scenarios via our augmentations. The synthetic pilots created aid in performance for re-training of digital receivers, while incurring only small overhead in computations (i.e. small number of additions and multiplications). Our proposed augmentations are not limited to the specific use-case, and can be further explored for other deep-learning aided communication tasks.   
	\vspace{-0.2cm}
	\bibliographystyle{IEEEtran}
	\bibliography{IEEEabrv,refs}
\end{document}